\documentclass{article}

\pdfoutput=1

\usepackage{graphicx} 
\usepackage{color}
\usepackage{amsmath} 
\usepackage{amssymb}
\usepackage{amsfonts}
\usepackage[margin=1.25in]{geometry}
\usepackage{rotating}
\usepackage{multirow}

\newcommand{\ch}[1]{\multicolumn{1}{c}{#1}}
\newcommand{\rh}[1]{\multirow{12}{1em}{\rotatebox{90}{#1}}}
\newcommand{\rhv}[1]{\multirow{6}{1em}{\rotatebox{90}{#1}}}

\title{Weak Scaling of DSA Preconditioning of Transport Sweeps using HYPRE} 

\author{Milan \textsc{Hanu{\v s}}} 

\date{February 8, 2017} 

\begin{document}

\maketitle 


\section{Introduction}

This report summarizes the weak scaling performance of the diffusion-synthetic acceleration (DSA)
of transport sweeps in PDT (a massively parallel deterministic transport solver developed at Texas A\&M University). The goal is to assess the cost of the DSA based on the symmetric interior-penalty discontinuous Galerkin discretization and solved by the AMG-preconditioned conjugate gradient method provided by the HYPRE/BoomerAMG library (which we will refer to shortly as HYPRE in the following), relative to the  transport sweeps. 

The following section gives an overview of the problem that has been solved and the methods used. Section \ref{sec:exp-hyb} presents the results obtained when using the automatic parallel partitioning of PDT suited for transport sweeps. This section is divided into two main subsections, which serve as a comparison of the HYPRE performance with parameter setup deemed as ``reasonable defaults'' in PDT, against the empirically obtained parameters based on other runs and several recent papers on AMG parallelization. Section \ref{sec:vol} presents the results for the volumetric partitioning, which is expected to perform better for the diffusion solves.

\section{Problem Description and Solution Methodology}

We solved a 1-group transport problem in a homogeneous cube of side length 16 cm, bounded from all sides by vacuum, with unit isotropic source and scattering ratio $c = 0.9999$. The piecewise linear discontinuous finite element approximation was used to discretize both the transport and the diffusion operators in space. The problem was solved in two workload regimes: 
\begin{itemize}
\item 512 cells per core $\longrightarrow$ $4096\times M$ degrees of freedom per core
\item 4096 cells per core $\longrightarrow$ $32768\times M$ degrees of freedom per core
\end{itemize}
where $M=1$ for the DSA problem, while for the transport problem $M$ is the number of angles in an angle set that aggregates a portion of the total number of directions used to discretize the angular domain. To this end, the product Gauss-Legendre-Chebyshev quadrature was used, with the following three resolutions:
\begin{itemize}
\item 4 polar, 4 azimuthal angles $\longrightarrow$ $128$ total angles aggregated into 8 angle sets ($M = 16$)
\item 8 polar, 8 azimuthal angles $\longrightarrow$ $512$ total angles aggregated into 8 angle sets ($M = 64$)
\item 16 polar, 16 azimuthal angles $\longrightarrow$ $2048$ total angles aggregated into 16 angle sets ($M = 128$)
\end{itemize}
The motivation behind increasing the number of angles was to find the value for which the DSA time would become sufficiently negligible with respect to the total solve time. 

As the purpose of the calculations was to measure performance rather than obtain converged results, a fixed number of 10 source iterations (all-direction sweeps), as well as 10 PCG iterations per source iterations have been performed. We however kept track of the final residual for both transport and PCG to make sure we were always getting the solution of more or less the same level of accuracy (see Sec. \ref{sec:hyb-opt} where we needed to switch to residual based convergence for PCG for this reason).

The weak scaling runs were performed on the VULCAN machine for core counts ranging from 1 to 128k, using Hypre 2.10.0.


\section{Hybrid KBA-volumetric partitioning}\label{sec:exp-hyb}

The first set of results was obtained when using the automatic parallel partitioning provided by PDT (including the automatic aggregation of cells to cell-sets and angles to angle-sets). This setting has been shown to perform well for the transport sweeps, but here we will see that it is not so well suited for the preconditioned diffusion solves. Table \ref{tab:part} shows the partitioning for the used core counts.

\begin{table}[htb]
\centering
\begin{tabular}{r|rrr}
\#cores & $P_x$ & $P_y$ & $P_z$ \\ \hline
1       & 1     & 1     & 1     \\
8       & 2     & 2     & 2     \\
64      & 1     & 8     & 8     \\
512     & 2     & 16    & 16    \\
1024    & 16    & 32    & 2     \\
2048    & 2     & 32    & 32    \\
4096    & 1     & 64    & 64    \\
8192    & 64    & 64    & 2     \\
16384   & 2     & 64    & 128   \\
32768   & 2     & 128   & 128   \\
65536   & 128   & 256   & 2     \\
131072  & 2     & 256   & 256  
\end{tabular}
\caption{Automatic parallel partitioning by PDT (hybrid-KBA).}
\label{tab:part}
\end{table}

\subsection{PDT-default HYPRE parameters}\label{sec:def-param}

Table \ref{tab:def} summarizes the default settings of HYPRE parameters that PDT uses for 3D diffusion problems. They are mostly based on the recommended values from HYPRE manual.

\begin{figure}[t]
\centering
\includegraphics[scale=.7]{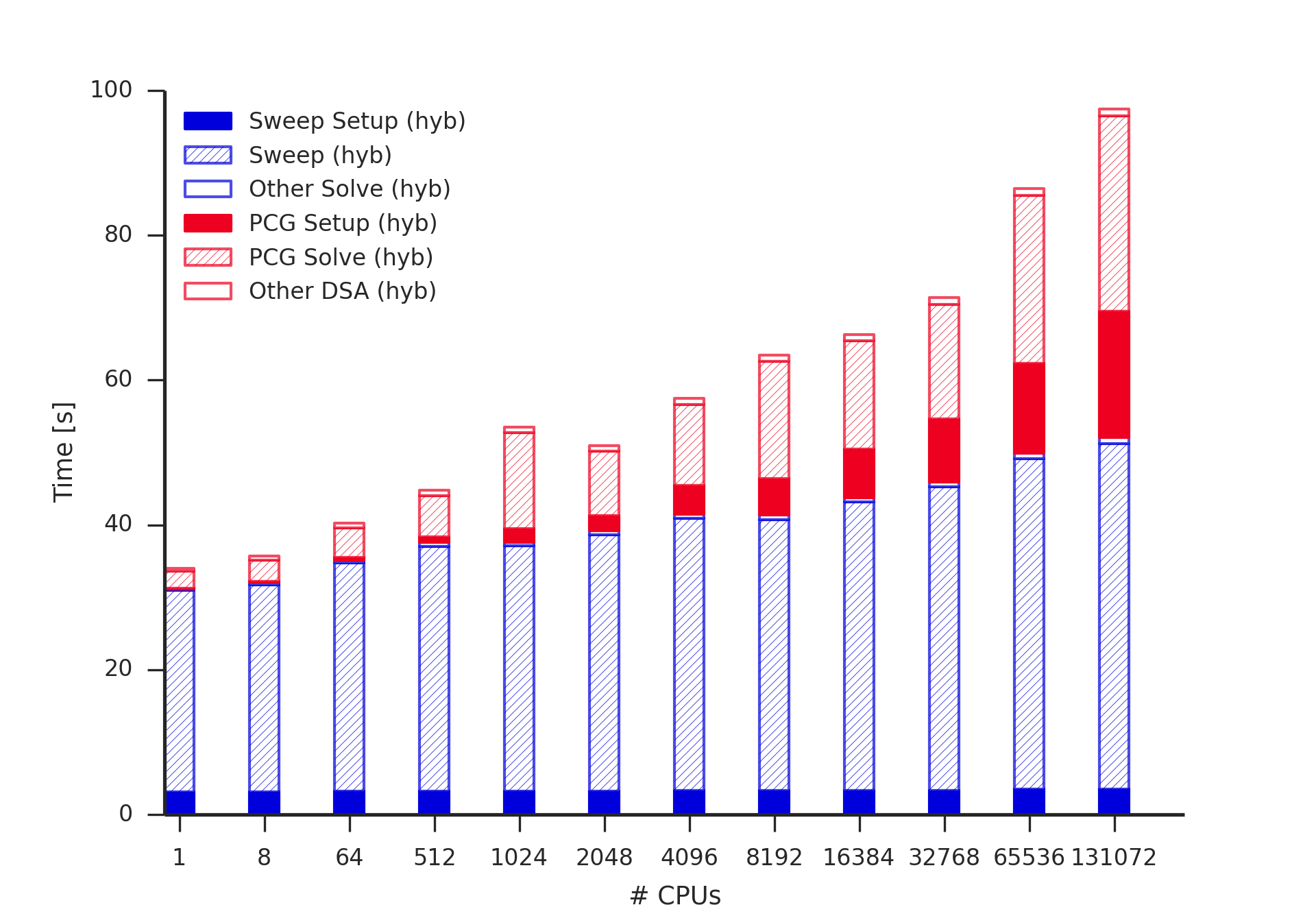}
\caption{Solve times, 512 cells/core, 512 directions}
\label{fig:default-512_512}
\end{figure}

\begin{figure}[b]
\centering
\includegraphics[scale=.7]{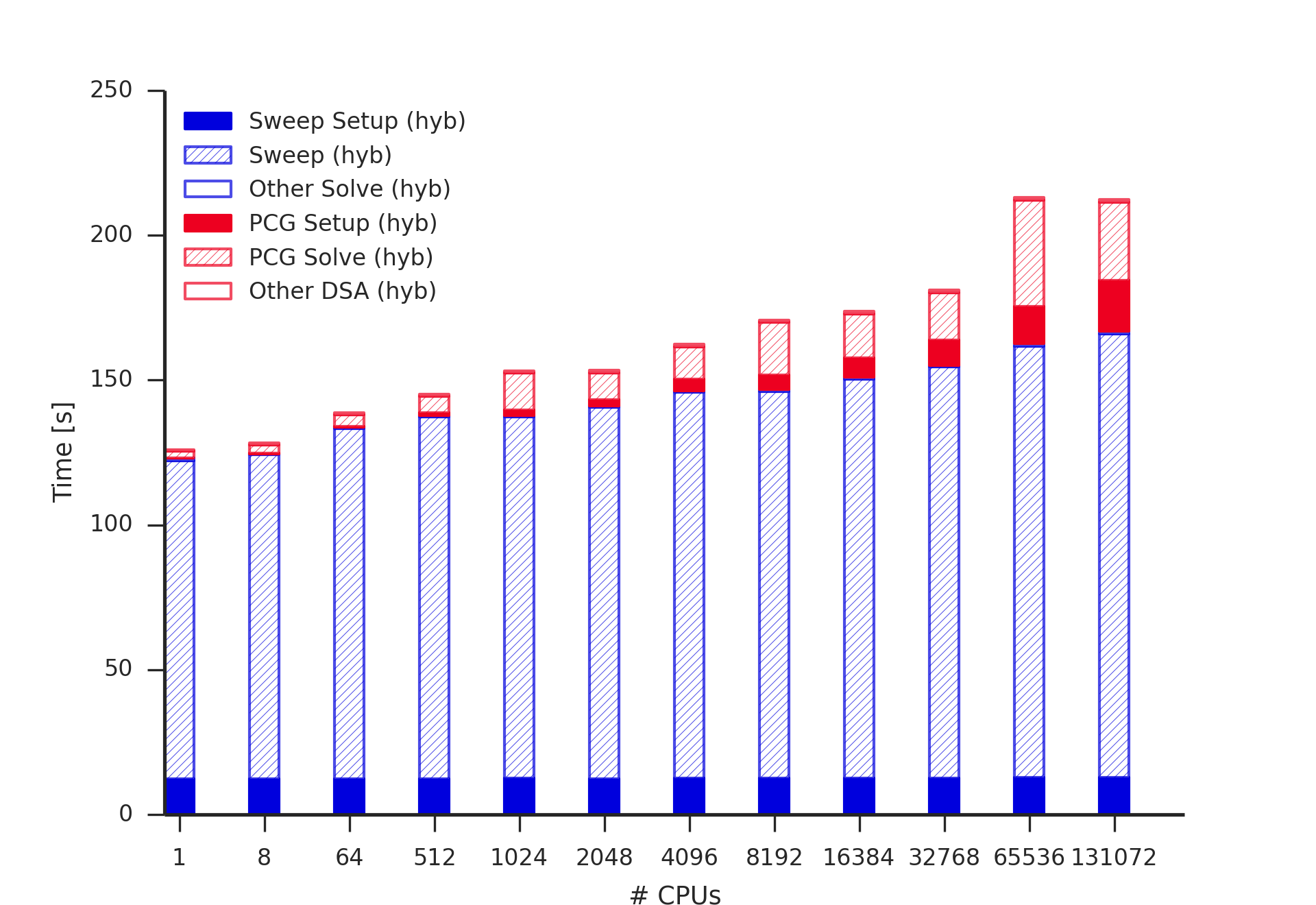}
\caption{Solve times, 512 cells/core, 2048 directions}
\label{fig:default-512_2048}
\end{figure}

\begin{figure}[t]
\centering
\includegraphics[scale=.7]{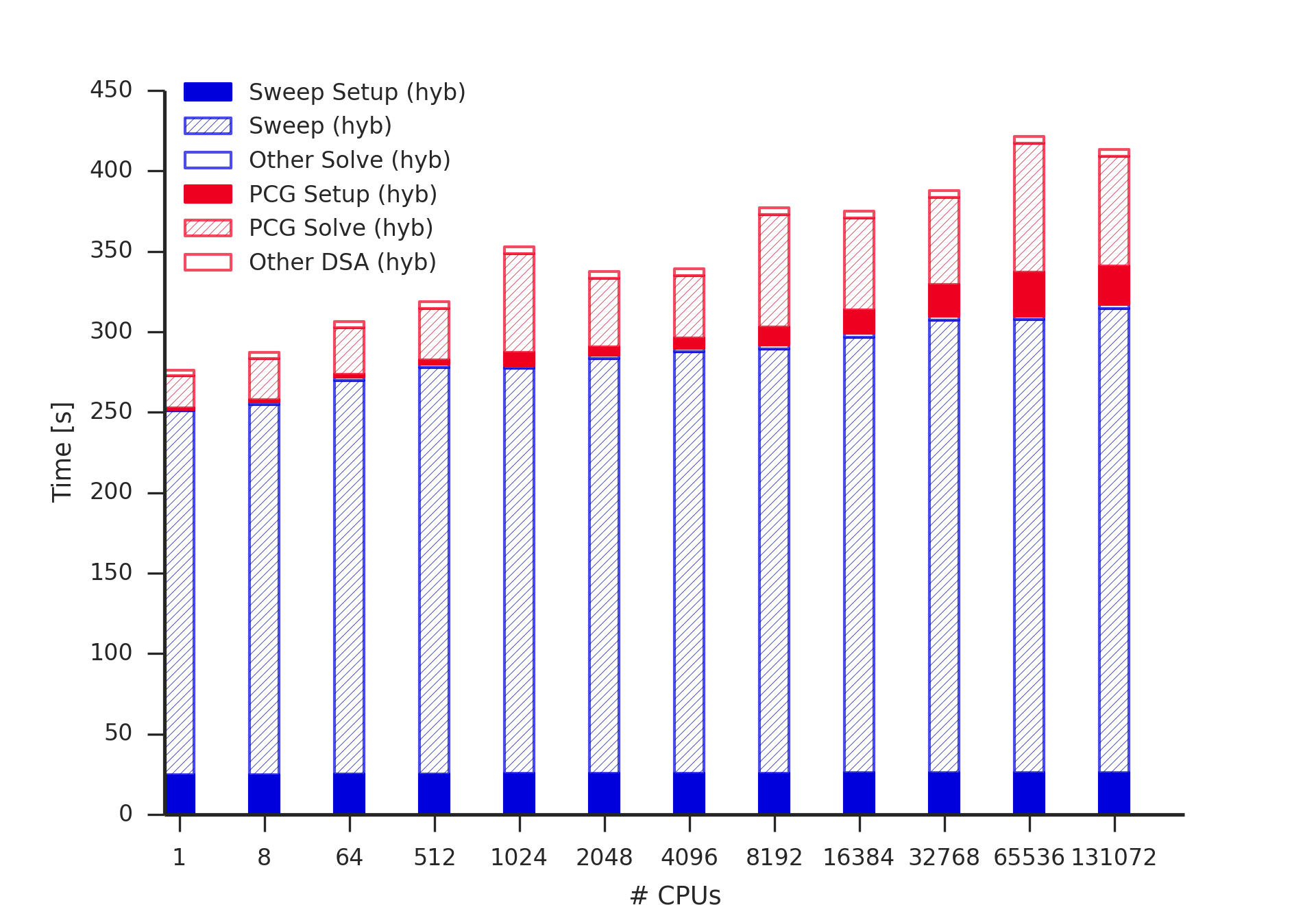}
\caption{Solve times, 4096 cells/core, 512 directions}
\label{fig:default-4096_512}
\end{figure}

\begin{figure}[b]
\centering
\includegraphics[scale=.7]{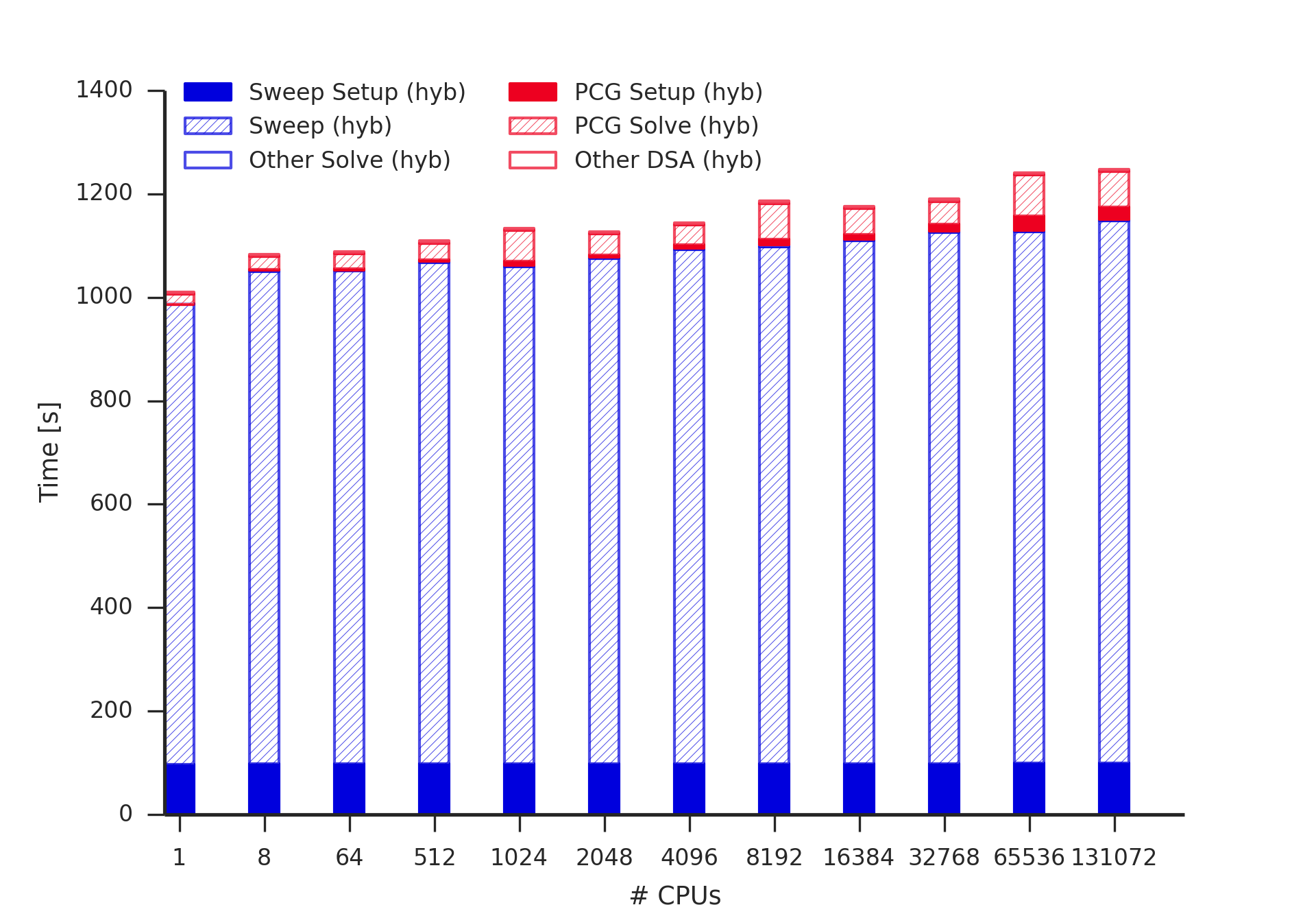}
\caption{Solve times, 4096 cells/core, 2048 directions}
\label{fig:default-4096_2048}
\end{figure}

Figures \ref{fig:default-512_512} through \ref{fig:default-4096_2048} show the total solve time decomposed by its main components:
\begin{itemize}
\item \textbf{PCG Setup} contains the construction of the AMG preconditioner and the setup of the HYPRE PCG solver (essentially calls to \texttt{HYPRE\_BoomerAMGSetup} and \linebreak[4]\texttt{HYPRE\_ParCSRPCGSetup}, respectively). The DSA matrix does not change between source iterations in a 1-group problem, which allows us to reuse the preconditioner; hence, PCG setup occurs only in the first iteration
\item \textbf{PCG Solve} measures the time spent by calling \texttt{HYPRE\_ParCSRPCGSolve}.
\item \textbf{Other DSA} contains all the remaining time spent in the DSA preconditioning phase, including the matrix/vector assembly and updates of the scalar flux.
\item \textbf{Sweep Setup} contains the preparation steps for the sweep, most importantly the construction of the sweep task graphs.
\item \textbf{Sweep} is the actual sweep for the whole problem domain (execution of the sweep task graphs).
\item \textbf{Other Solve} contain the remaining operations performed in the solve phase (construction of the source, computation of residuals, convergence checking, etc.).
\end{itemize}
All times were measured using internal PDT timing routines based on \texttt{gettimeofday()} and are accumulated over all iterations. Note that all the time components add to the \textit{total solve time}, which is different from the \textit{overall time} reported in the raw tables in the Appendix (this latter time includes also the grid construction and problem setup, which is the same for both the transport and the diffusion parts and hence was omitted from the charts below).

Figures \ref{fig:default-512_512} through \ref{fig:default-4096_2048} confirm good weak scalability of PDT sweeps, but reveal bad scalability of the diffusion solver (both the preconditioner and solve phase), especially with lower work loads. Poor scalability of AMG with Galerkin coarsening has been recently addressed in several papers (e.g., \cite{Bienz2016ReducingPC,Falgout2014NonGalerkinCG,precond,Gahvari}) so as an attempt to improve our DSA performance, we used the knowledge from these papers to experiment with the AMG parameters a bit. Although these experiments were performed earlier using a different test case (a multigroup graphite block problem), we used the final set of parameters that yielded best results for that problem here as well. We call this set ``optimized''. Next section shows the results obtained with these parameters for the present problem.

\begin{table}[!hbt]
\centering
\begin{tabular}{l|lp{.65\textwidth}}
Parameter      & Value & Brief description (see \cite{hypre} for more details)                                                                 \\ \hline
coarsening     & 10    & HMIS-coarsening                                                                                                             \\
relaxation     & 6     & hybrid symmetric Gauss-Seidel                                                                                               \\
interpolation  & 6     & extended+i interpolation                                                                                                    \\
strength thr   & 0.25  & AMG strength threshold                                                                                                      \\
num sweeps     & 1     & number of sweeps                                                                                                            \\
pmax elmts     & 4     & maximal number of elements per row for the interpolation                                                                    \\
agg levels     & 1     & number of levels of aggressive coarsening                                                                                   \\
agg interp     & 4     & interpolation used on levels of aggressive coarsening\\ 
& & (4 = multipass)                                                       \\
agg pmax el    & 0     & maximal number of elements per row for the interpolation used for aggressive coarsening (0 = unlimited)                     \\
agg p12 max el & 0     & maximal number of elements per row for the matrices P1 and P2 which are used to build 2-stage interpolation\\ 
& & (0 = unlimited) \\
amg nongalerk  & []    & list of non-Galerkin drop-tolerances at each level for sparsifying coarse grid operators and thus reducing communication\\
& & (empty = standard Galerkin coarsening at all levels)
\end{tabular}
\caption{Default HYPRE parameters in PDT.}
\label{tab:def}
\end{table}

\clearpage

\clearpage

\subsection{``Optimized'' HYPRE parameters}\label{sec:hyb-opt}

\begin{table}[hb]
\centering
\begin{tabular}{l|lp{.65\textwidth}}
Parameter      & Value & Brief description (see \cite{hypre} for more details)                                                                 \\ \hline
coarsening     & 10    & HMIS-coarsening                                                                                                             \\
relaxation     & 6     & hybrid symmetric Gauss-Seidel                                                                                               \\
interpolation  & 6     & extended+i interpolation                                                                                                    \\
strength thr   & \textbf{0.8}  & AMG strength threshold                                                                                                      \\
num sweeps     & 1     & number of sweeps                                                                                                            \\
pmax elmts     & 4     & maximal number of elements per row for the interpolation                                                                    \\
agg levels     & \textbf{10}     & number of levels of aggressive coarsening                                                                                   \\
agg interp     & 4     & interpolation used on levels of aggressive coarsening\\ 
& & (4 = multipass)                                                       \\
agg pmax el    & \textbf{4}     & maximal number of elements per row for the interpolation used for aggressive coarsening (0 = unlimited)                     \\
agg p12 max el & 0     & maximal number of elements per row for the matrices P1 and P2 which are used to build 2-stage interpolation\\ 
& & (0 = unlimited) \\
amg nongalerk  & \textbf{[0.0 0.1]}    & list of non-Galerkin drop-tolerances at each level for sparsifying coarse grid operators and thus reducing communication (0.0 at the finest level, 0.1 on all others)
\end{tabular}
\caption{``Optimized'' HYPRE parameters (changes from the defaults in bold).}
\label{tab:opt}
\end{table}

\begin{figure}[!ht]
\centering
\includegraphics[scale=.7]{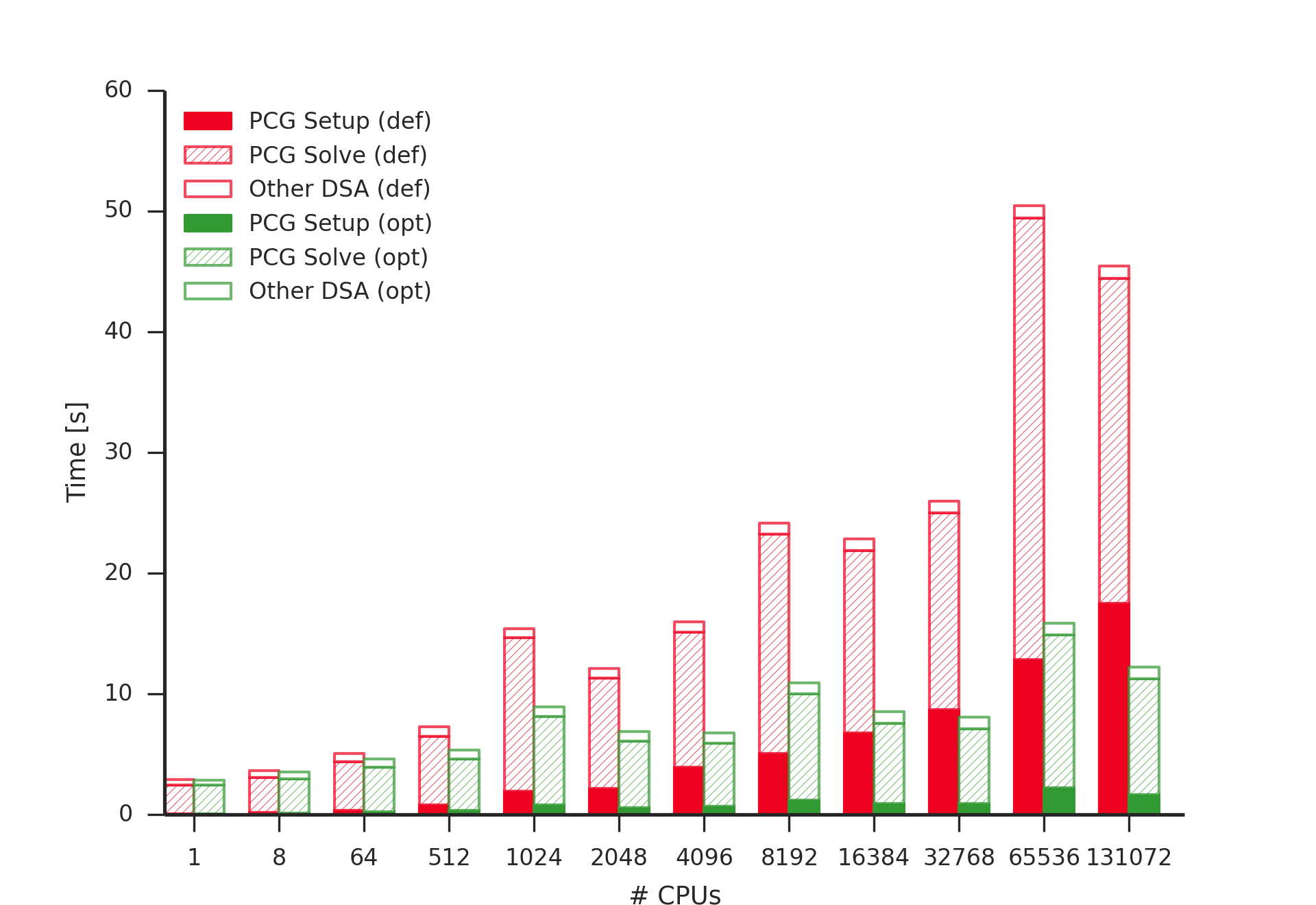}
\caption{HYPRE times, 512 cells/core, 2048 directions, PCG max\_it = 10.}
\label{fig:default-512_2048-vs-opt_fixed-512_2048}
\end{figure}

\begin{figure}[!hb]
\centering
\includegraphics[scale=.7]{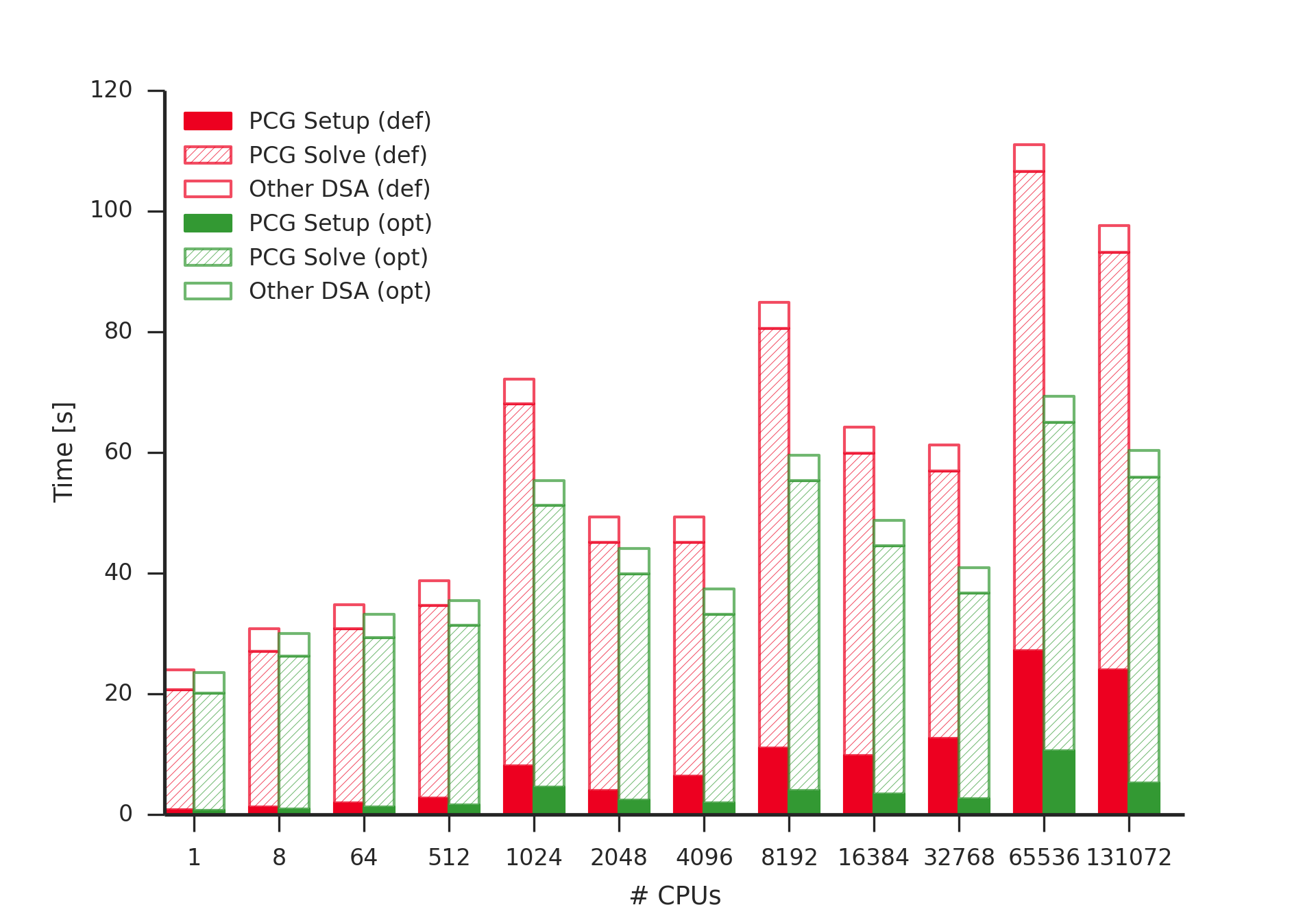}
\caption{HYPRE times, 4096 cells/core, 2048 directions, PCG max\_it = 10.}
\label{fig:default-4096_2048-vs-opt_fixed-4096_2048}
\end{figure}

\begin{figure}[htb]
\centering
\includegraphics[scale=.7]{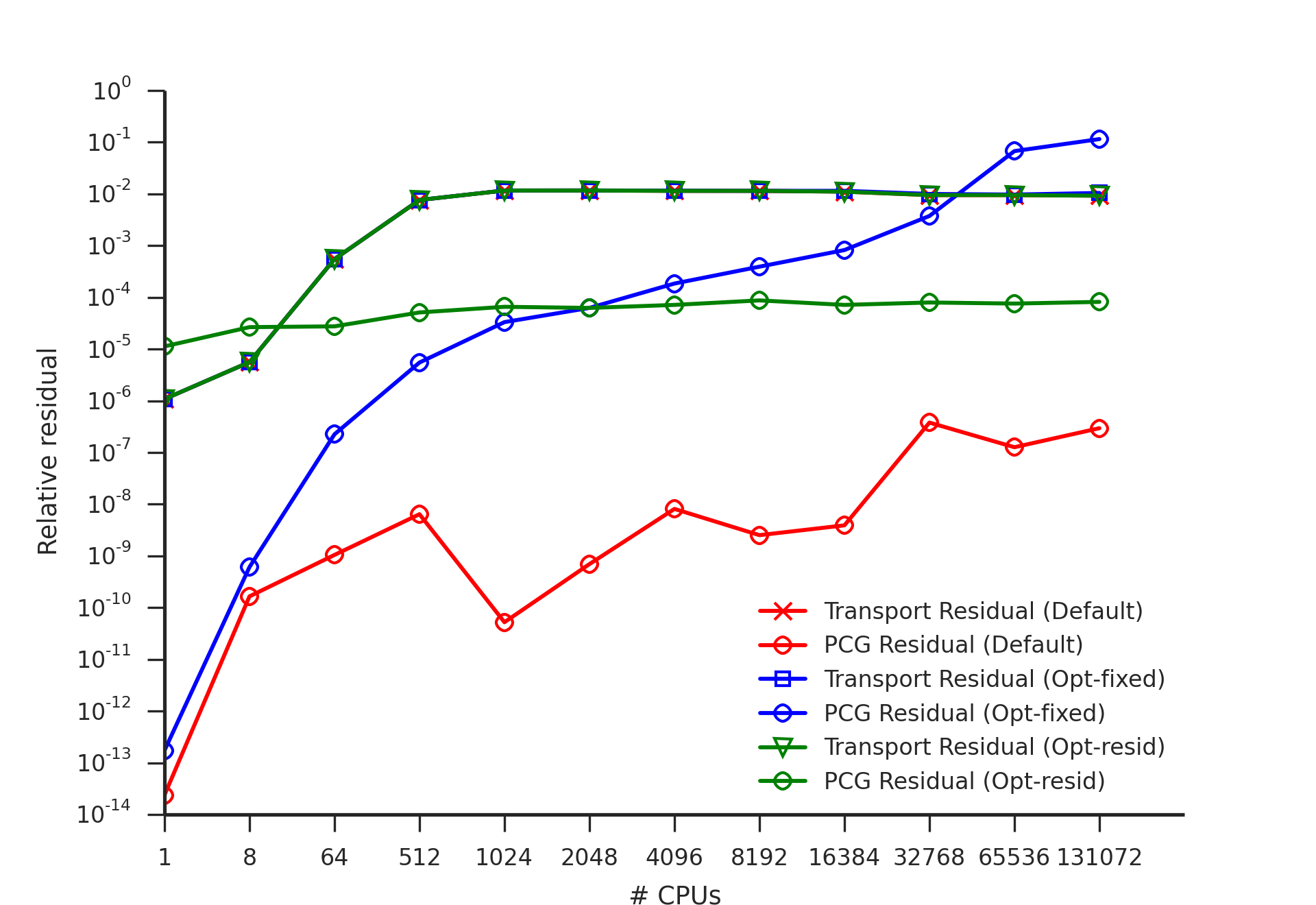}
\caption{Final residuals, 512 cells/core, 2048 directions.}
\label{fig:counters-512_2048-resid}
\end{figure}

\begin{figure}[htb]
\centering
\includegraphics[scale=.7]{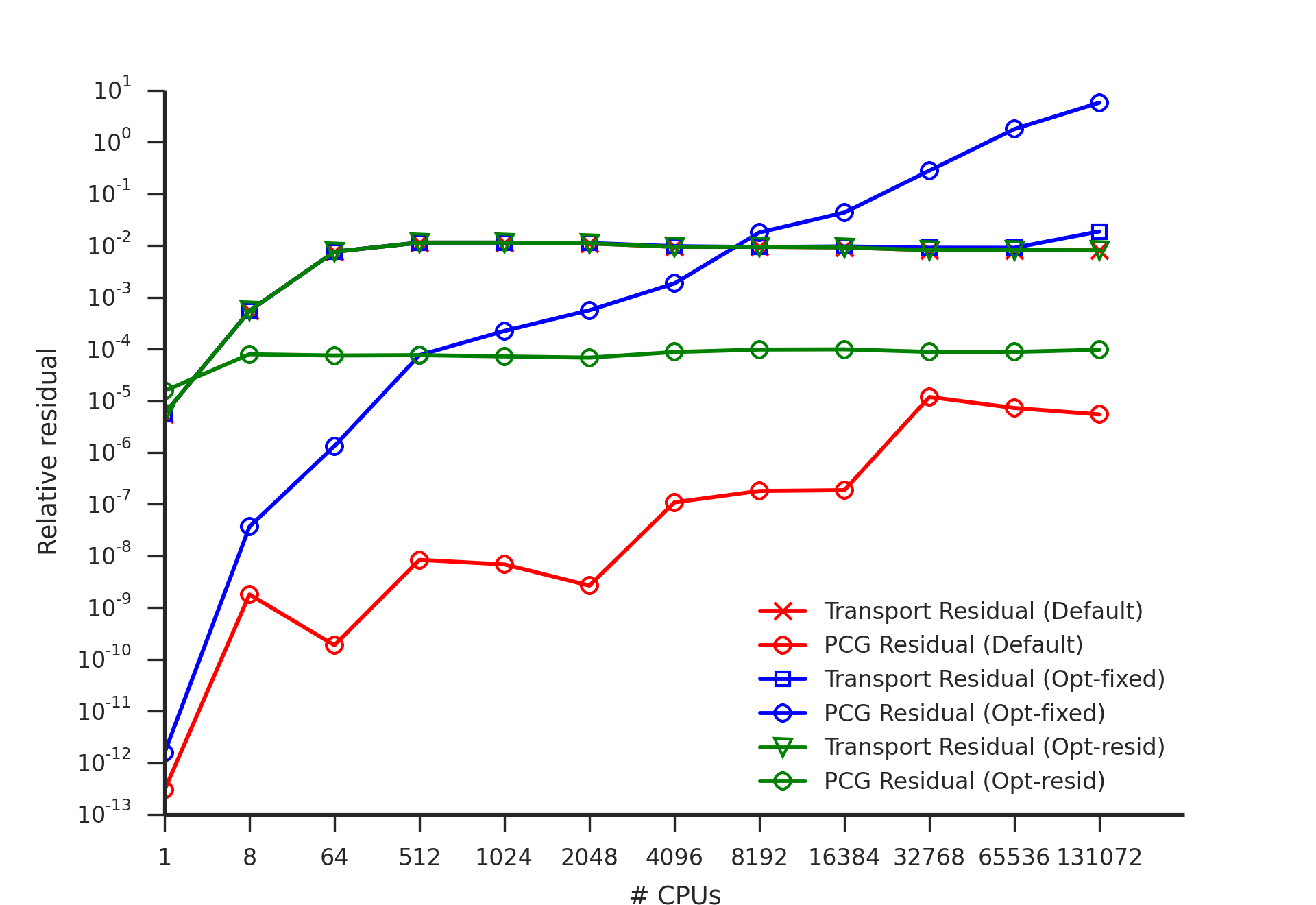}
\caption{Final residuals, 4096 cells/core, 2048 directions.}
\label{fig:counters-4096_2048-resid}
\end{figure}

Figures \ref{fig:default-512_2048-vs-opt_fixed-512_2048} and \ref{fig:default-4096_2048-vs-opt_fixed-4096_2048} compare the HYPRE times with the PDT-default settings and the optimized settings. While they show significant improvement for the AMG setup time as well as for the PCG solve time (at higher core counts), we must keep in mind that the price for the lower cost of AMG setup is that the preconditioner is not as effective. The red and blue curves in Figures \ref{fig:counters-512_2048-resid} and \ref{fig:counters-4096_2048-resid} show the transport and PCG residuals attained after the total of 10 source iterations and 100 PCG iterations (10 PCG iterations per source iteration). With increasing core counts (and decreasing cell sizes), the PCG residuals generally grow, and especially fast with the optimized settings. Nevertheless, this comparison with fixed number of PCG iterations indicates that there is a relatively big potential of reducing both the setup cost and solve cost (essentially the cost of matrix-vector multiplication) by playing with the parameters that HYPRE exposes to the user.

In the case of 4096 cells per core and 128k cores, the decrease of the effectivity of the preconditioner for the diffusion solves started to have a detrimental effect on the convergence of the transport solution itself  (see the blue line with square markers in Fig. \ref{fig:counters-4096_2048-resid}). Hence, in order to obtain fair practical comparison, we switched the convergence criterion for the DSA from fixed number of PCG iterations to a prescribed PCG residual of $10^{-4}$. The green lines in figures \ref{fig:counters-512_2048-resid} and \ref{fig:counters-4096_2048-resid} indicate that using this convergence criterion, the DSA with the cheaper AMG preconditioner setup is (for this particular problem) sufficient for attaining the same transport residual as DSA with the more expensive AMG obtained with the default PDT settings. However, the number of PCG iterations required to attain the prescribed PCG residual value naturally increases at higher core counts. 

\begin{figure}[htb]
\centering
\includegraphics[scale=.7]{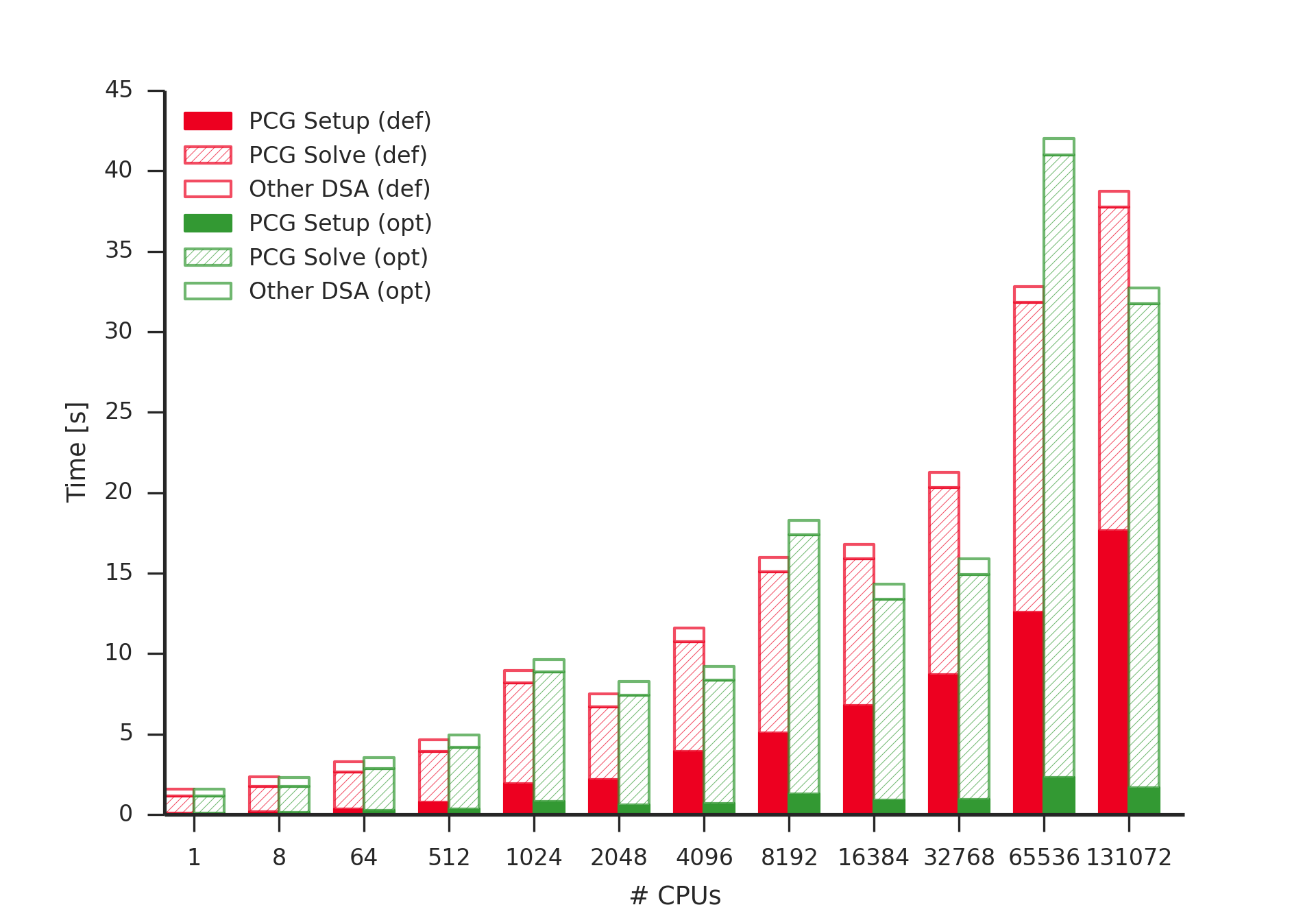}
\caption[]{%
\begin{tabular}[t]{l}%
  HYPRE times, 512 cells/core, 2048 directions,\\%
  PCG res\_thr = 10$^{-4}$.%
\end{tabular}}
\label{fig:default-512_2048-vs-opt_resid-512_2048}
\end{figure}

\begin{figure}[htb]
\centering
\includegraphics[scale=.7]{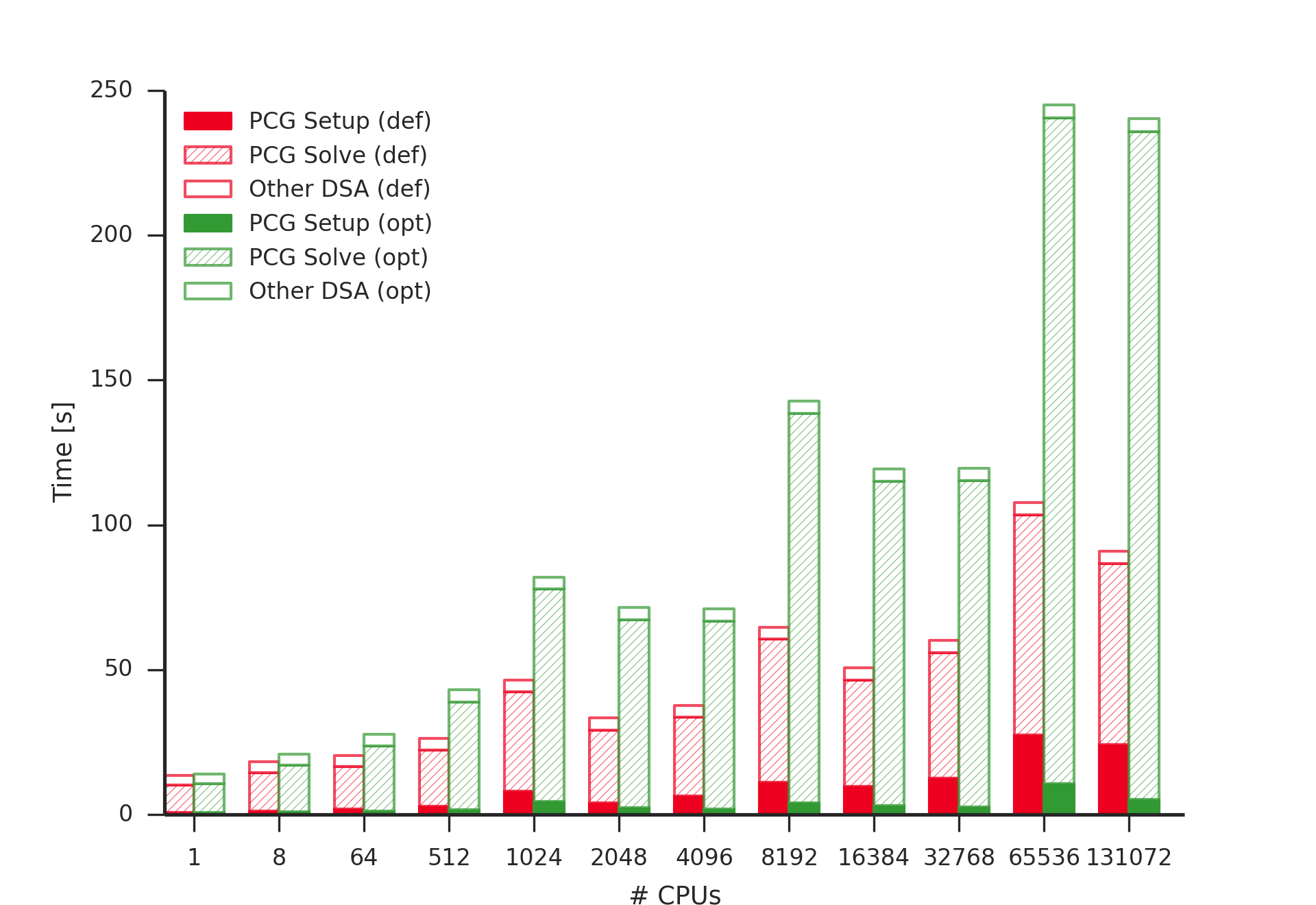}
\caption[]{%
\begin{tabular}[t]{l}%
  HYPRE times, 4096 cells/core, 2048 directions,\\%
  PCG res\_thr = 10$^{-4}$.%
\end{tabular}}
\label{fig:default-4096_2048-vs-opt_resid-4096_2048}
\end{figure}

Figures \ref{fig:default-512_2048-vs-opt_resid-512_2048} and \ref{fig:default-4096_2048-vs-opt_resid-4096_2048} illustrate the implications of this growth of PCG iterations -- while for 512 cells per core, the optimized settings still lead to slightly better DSA times at higher core counts\footnote{Some of the 65k-core runs resulted in spikes in PCG solve times (per iteration) that we currently cannot explain. See also the raw data tables in the Appendix.}, with 4096 cells per core, the increased number of PCG iterations clearly offsets any gains achieved in the AMG setup phase. This is likely a consequence of the way we obtained the ``optimized'' parameter set -- for the original multi-group problem, we started with the scheme where the AMG setup for the within-group DSA was performed for every group, thus the emphasis on minimizing the AMG setup times. There is most likely a different parameter set that would favor the PCG solve time (or strike better balance between both), better suited for one-group problems (or problems where the DSA for different groups can be performed using the same setup). However, we have not yet attempted to find such a set.

\begin{figure}[h]
\centering
\includegraphics[scale=.7]{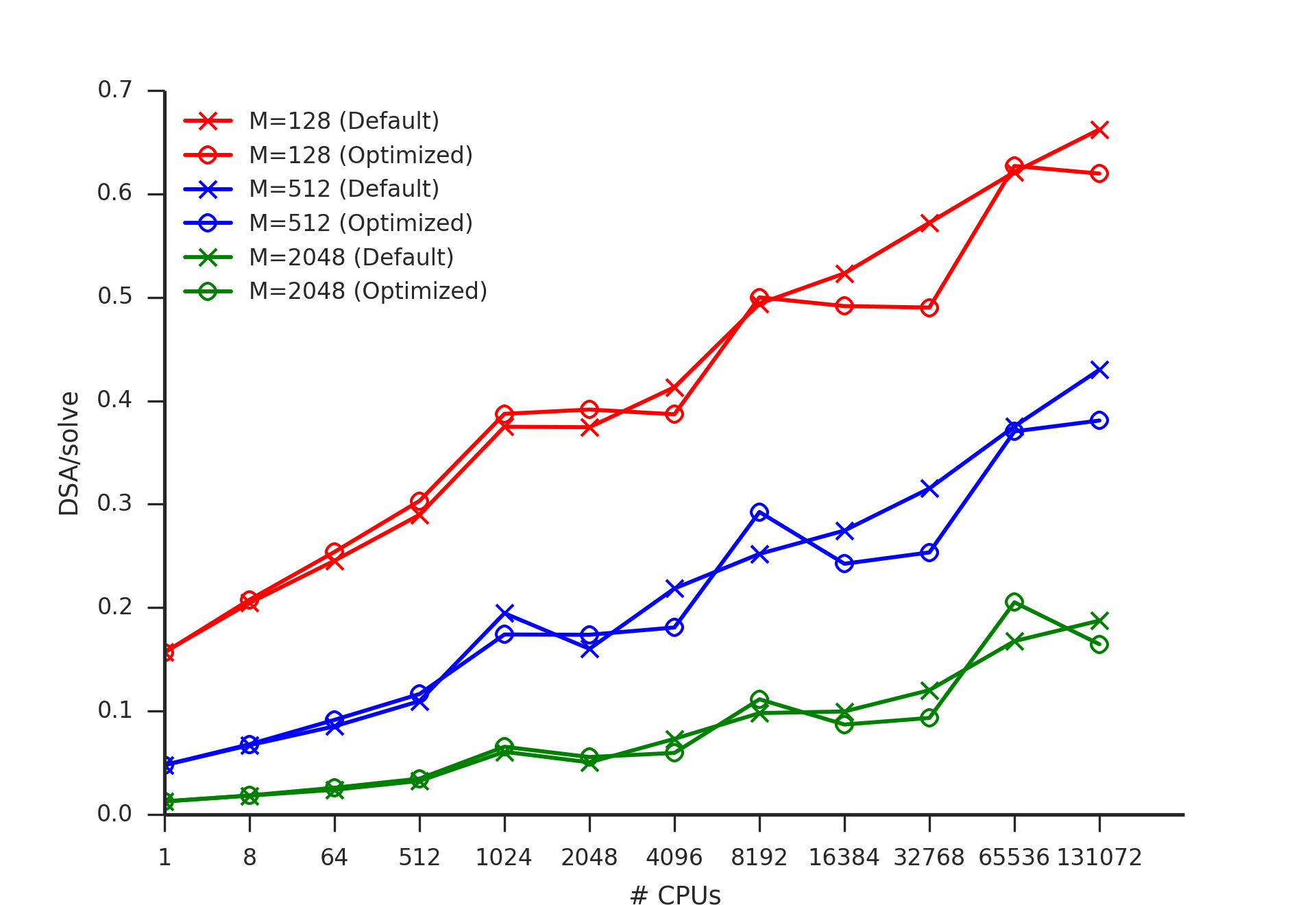}
\caption{DSA/solve fractions, 512 cells/core, PCG res\_thr = 10$^{-4}$}
\label{fig:dsa2solve-512}
\end{figure}

\begin{figure}[!hb]
\centering
\includegraphics[scale=.7]{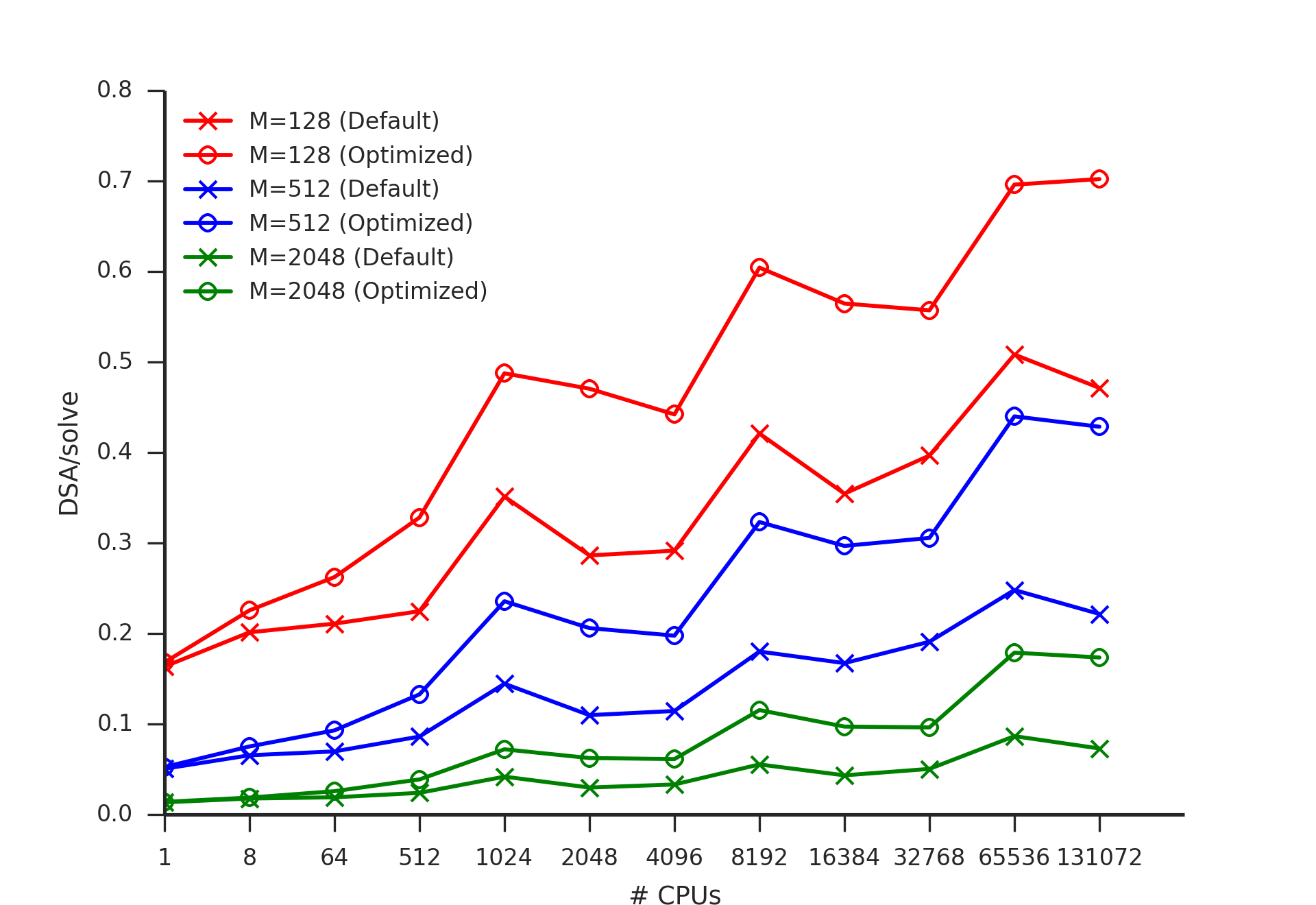}
\caption{DSA/solve fractions, 4096 cells/core, PCG res\_thr = 10$^{-4}$}
\label{fig:dsa2solve-4096}
\end{figure}

All these results are finally reflected in the fractions of DSA to the total solve time for the three sets of angular discretizations and the default and the optimized HYPRE settings (with PCG residual convergence criterion). These are plotted in figures \ref{fig:dsa2solve-512} and \ref{fig:dsa2solve-4096} for 512 cells per core and 4096 cells per core, respectively.

\clearpage 


\clearpage

\clearpage


\section{Volumetric partitioning}\label{sec:vol}

We solved the problem on 1, 8, 64, 512, 4096 and 32768 cores using the volumetric partitioning, creating a configuration that should be more favorable for the diffusion solver. We only tested the case with 2048 directions and empirically found the best angle aggregation for that configuration to be 8 angles per angle-set. 

\begin{figure}[!ht]
\centering
\includegraphics[width=.49\textwidth]{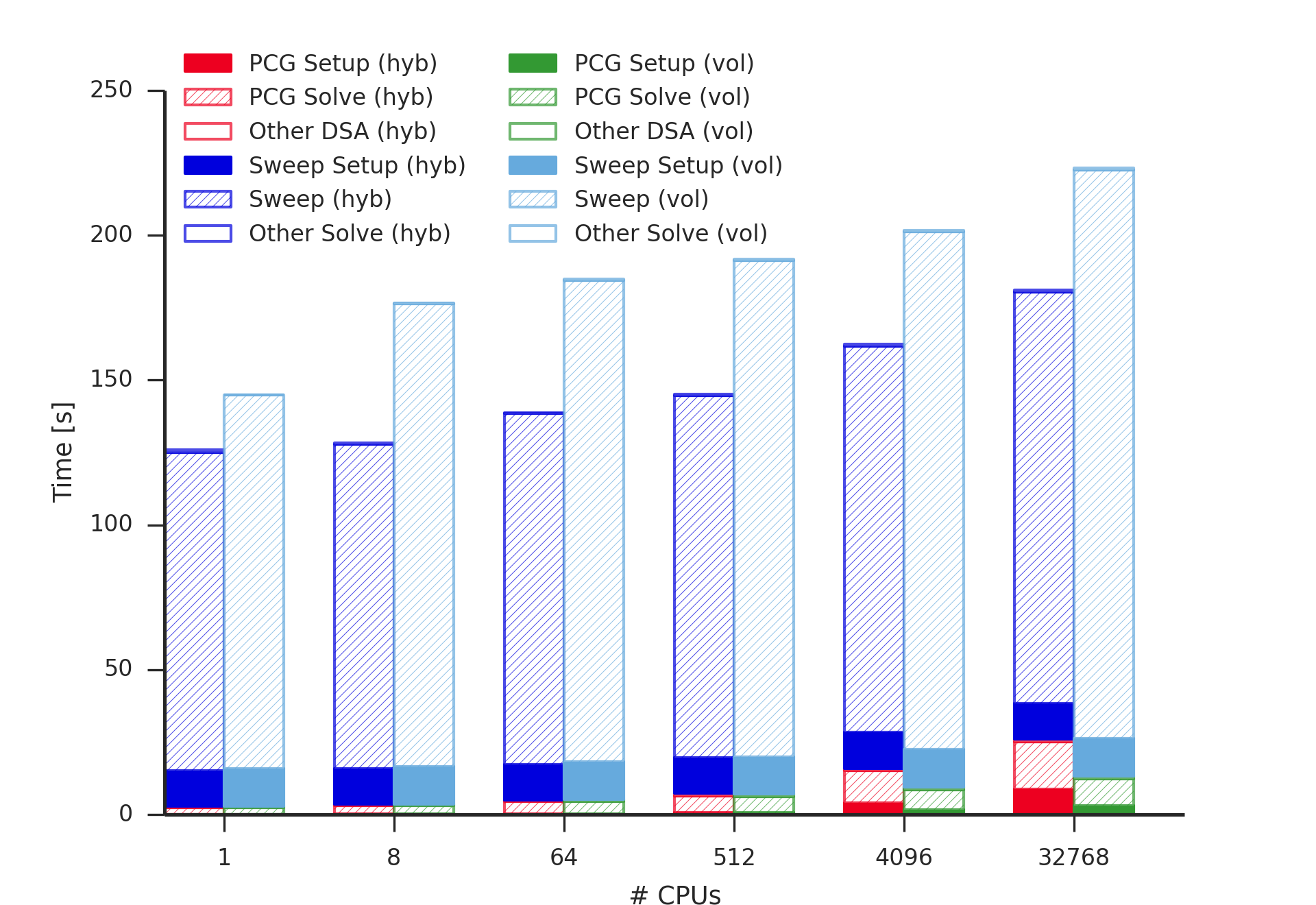}
\includegraphics[width=.49\textwidth]{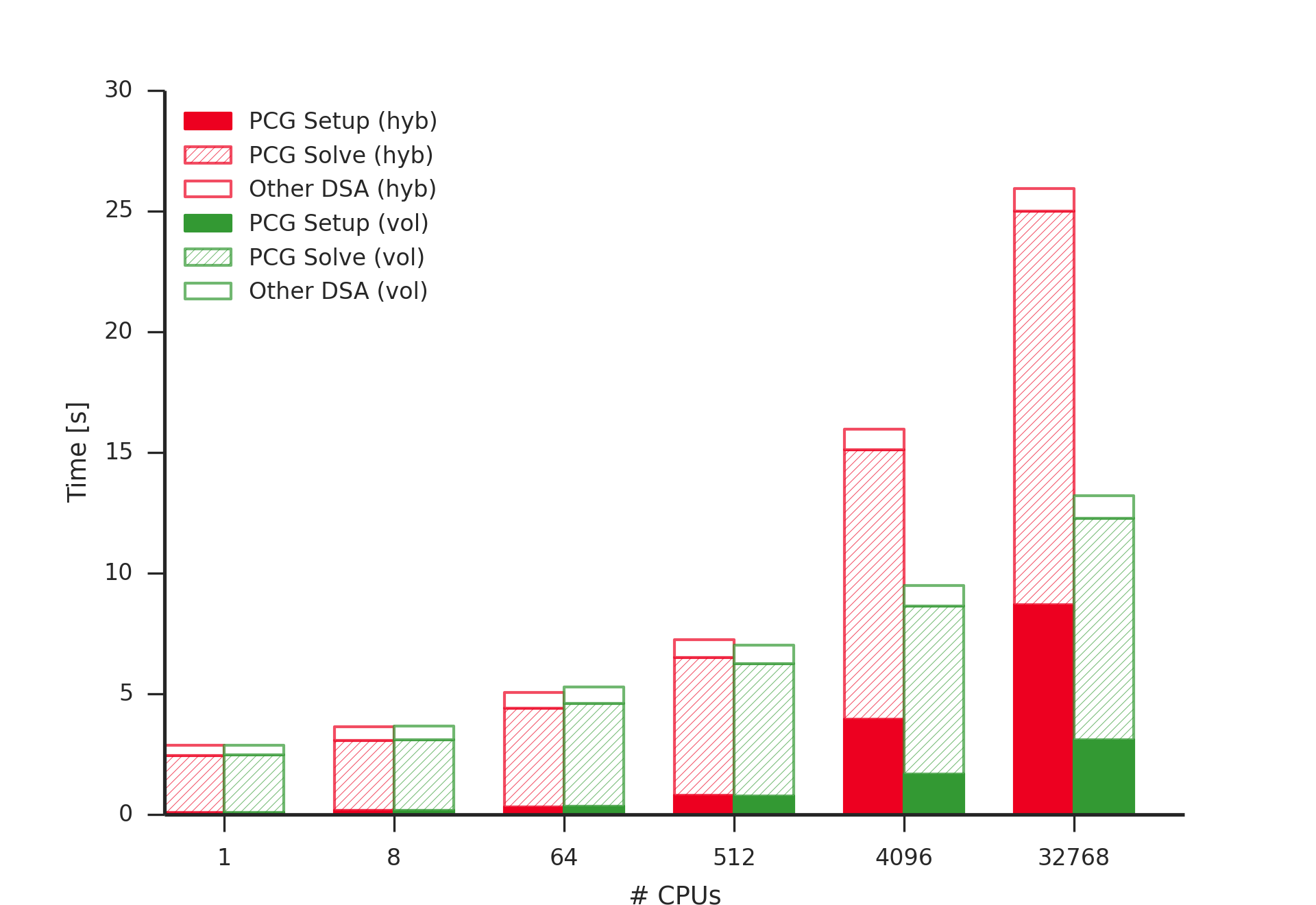}
\caption[]{\begin{tabular}[t]{l}%
  Hybrid-KBA vs. volumetric -- 512 cells/core, 2048 directions, default HYPRE parameters,\\%
  PCG max\_it = 10. \textit{Left}: all solve times, \textit{right}: HYPRE-only times.%
\end{tabular}}
\label{fig:default-512_2048-vs-default-512_2048_vol}
\end{figure}

\begin{figure}[!hb]
\centering
\includegraphics[width=.49\textwidth]{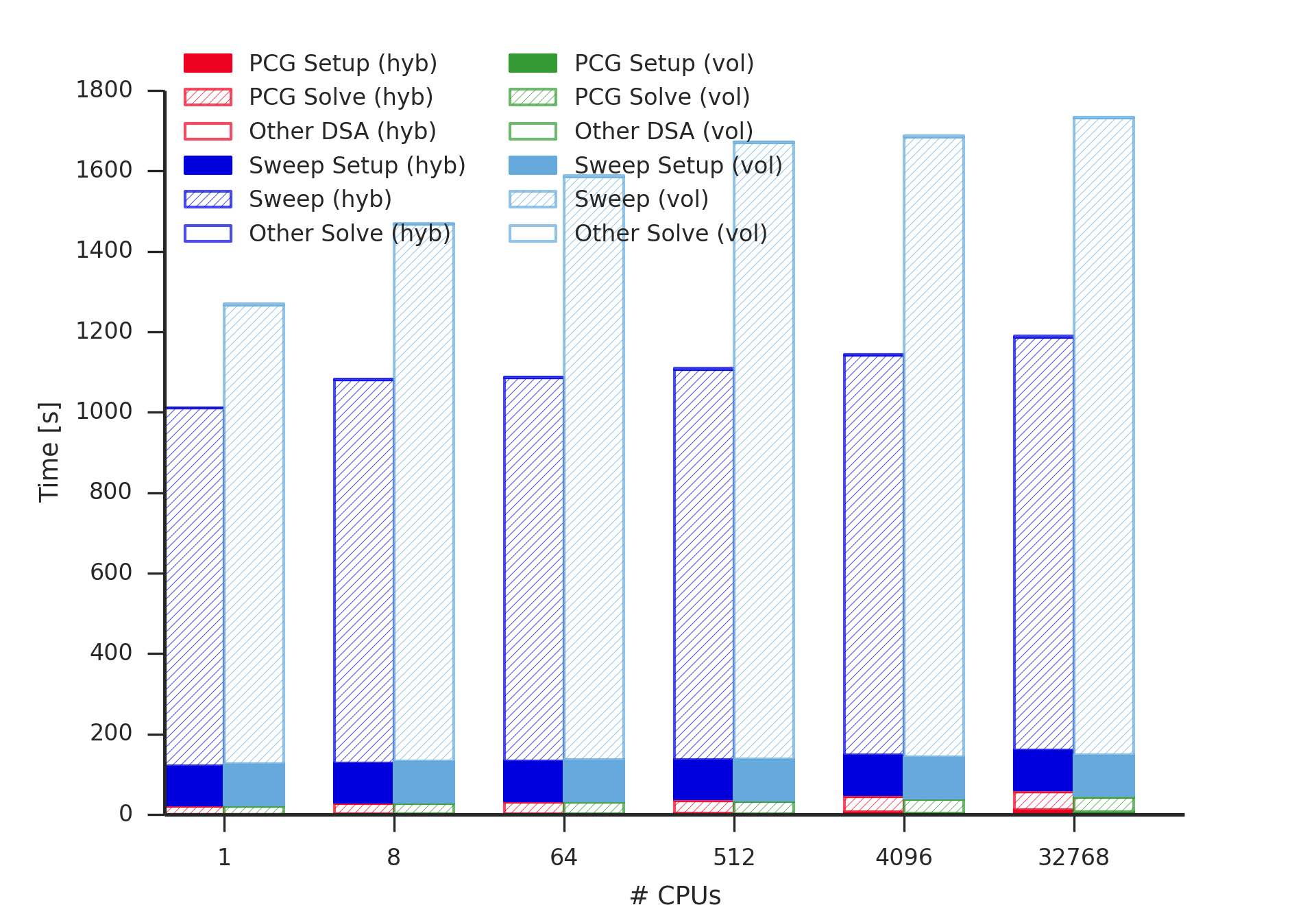}
\includegraphics[width=.49\textwidth]{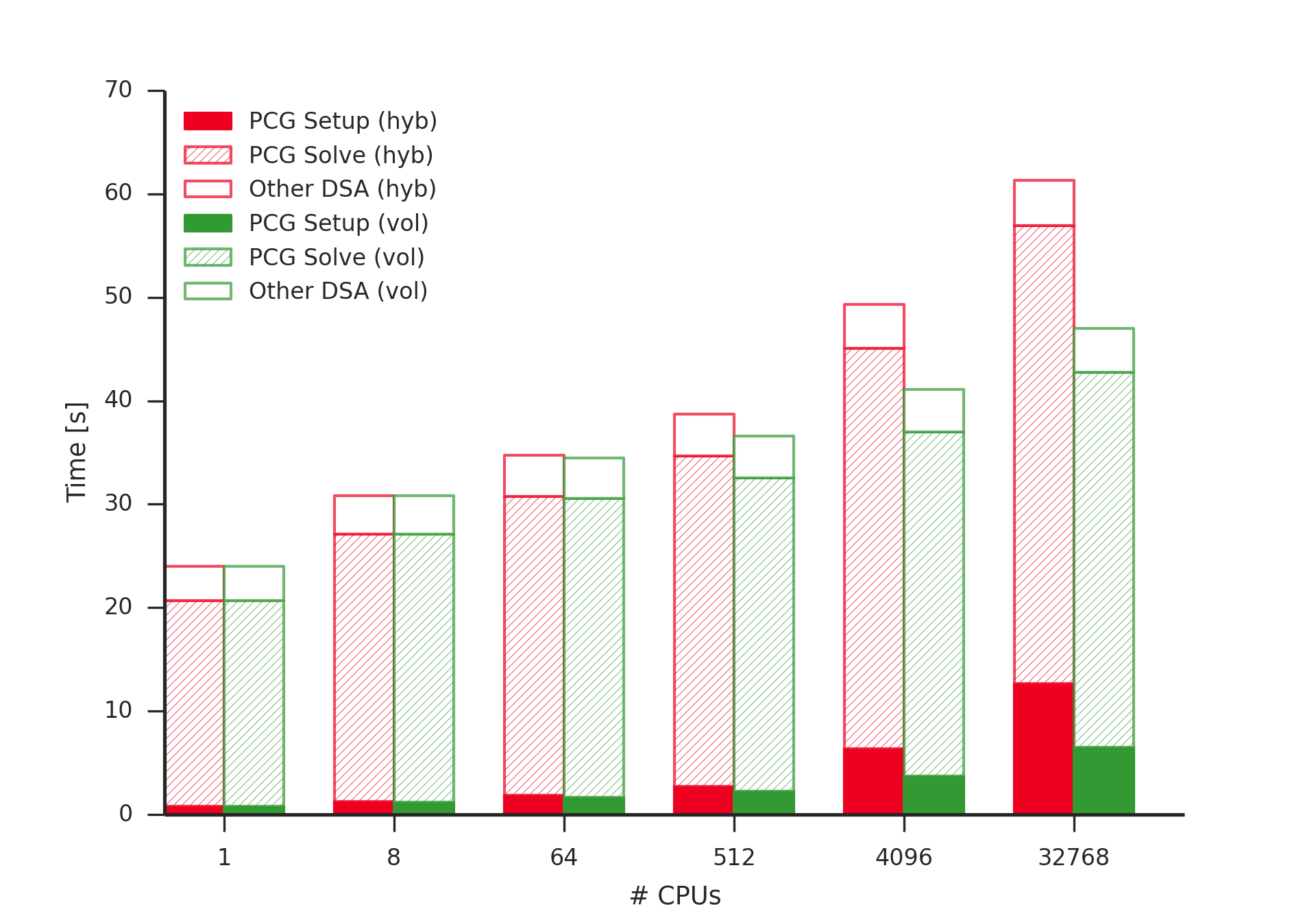}
\caption[]{\begin{tabular}[t]{l}%
  Hybrid-KBA vs. volumetric -- 4096 cells/core, 2048 directions, default HYPRE parameters,\\%
 PCG max\_it = 10. \textit{Left}: all solve times, \textit{right}: HYPRE-only times.%
\end{tabular}}
\label{fig:default-4096_2048-vs-default-4096_2048_vol}
\end{figure}

\begin{figure}[htb]
\centering.
\includegraphics[scale=.7]{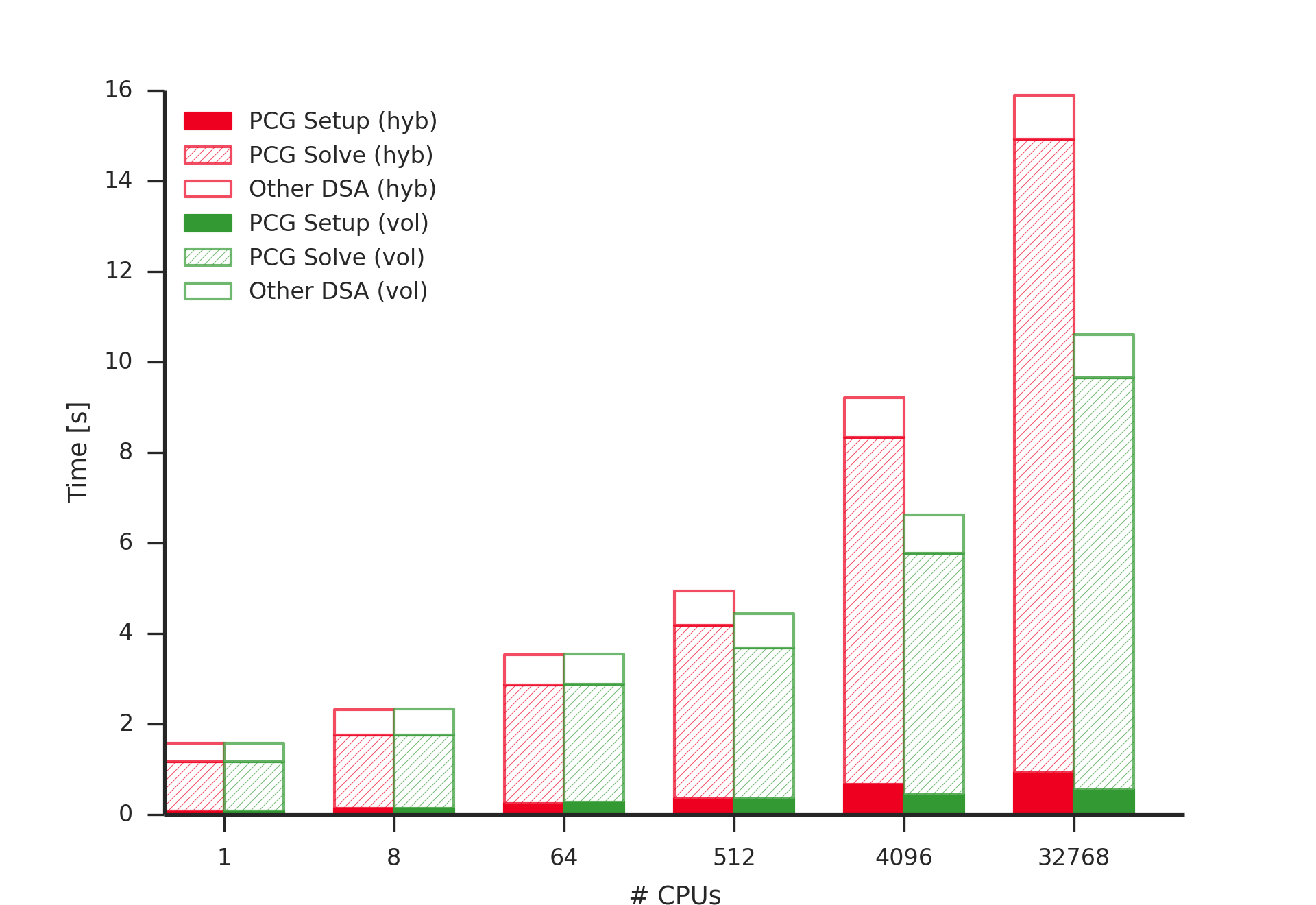}
\caption[]{\begin{tabular}[t]{l}%
  Hybrid-KBA vs. volumetric -- 512 cells/core, 2048 directions,\\%
  Optimized HYPRE parameters with PCG res\_thr = $10^{-4}$.%
\end{tabular}}
\label{fig:opt_resid-512_2048-vs-opt_resid-512_2048_vol}
\end{figure}

\begin{figure}[htb]
\centering
\includegraphics[scale=.7]{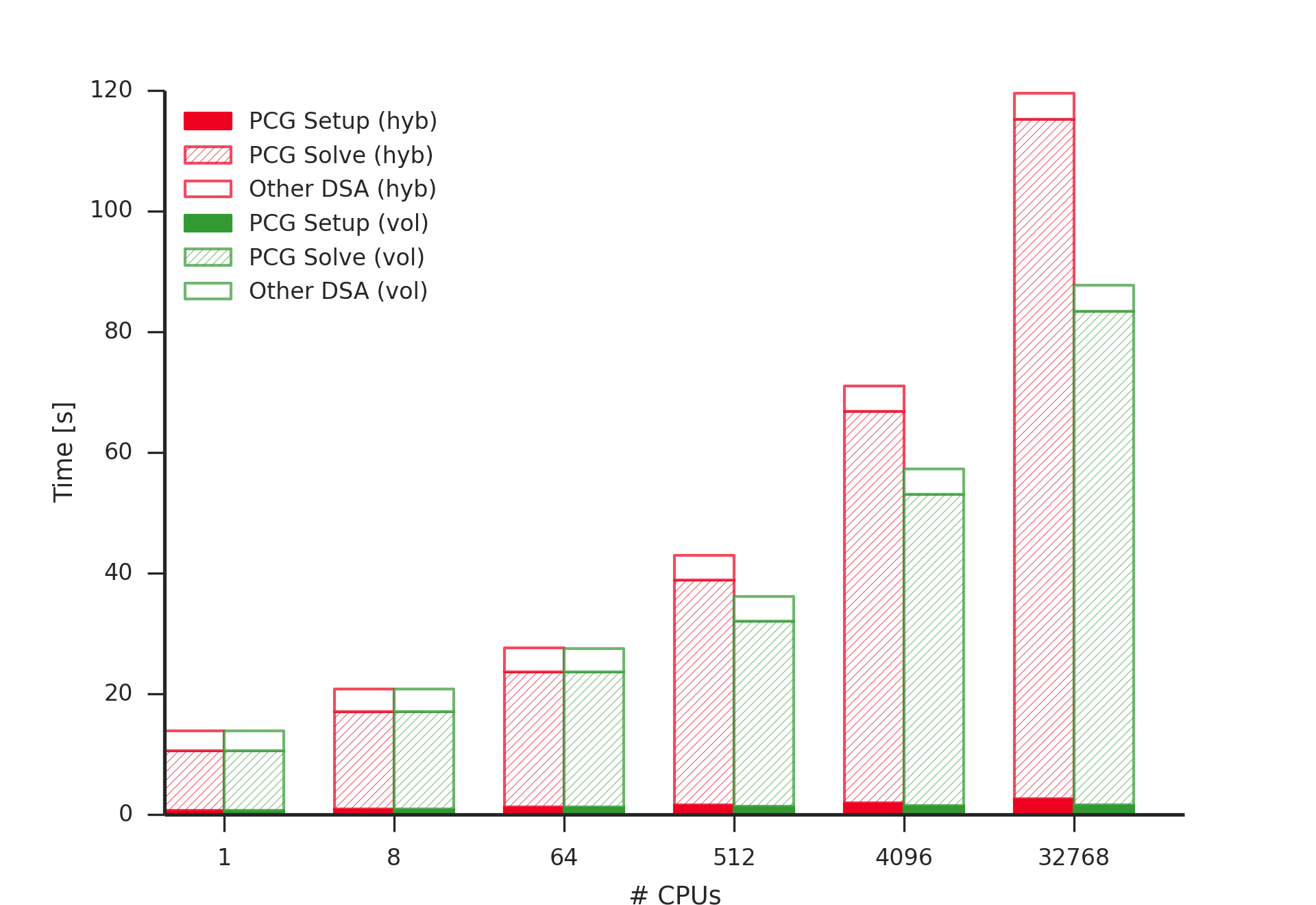}
\caption[]{\begin{tabular}[t]{l}%
  Hybrid-KBA vs. volumetric -- 4096 cells/core, 2048 directions,\\%
  Optimized HYPRE parameters with PCG res\_thr = $10^{-4}$.%
\end{tabular}}
\label{fig:opt_resid-4096_2048-vs-opt_resid-4096_2048_vol_hypre}
\end{figure}

\begin{figure}[htb]
\centering
\includegraphics[scale=.7]{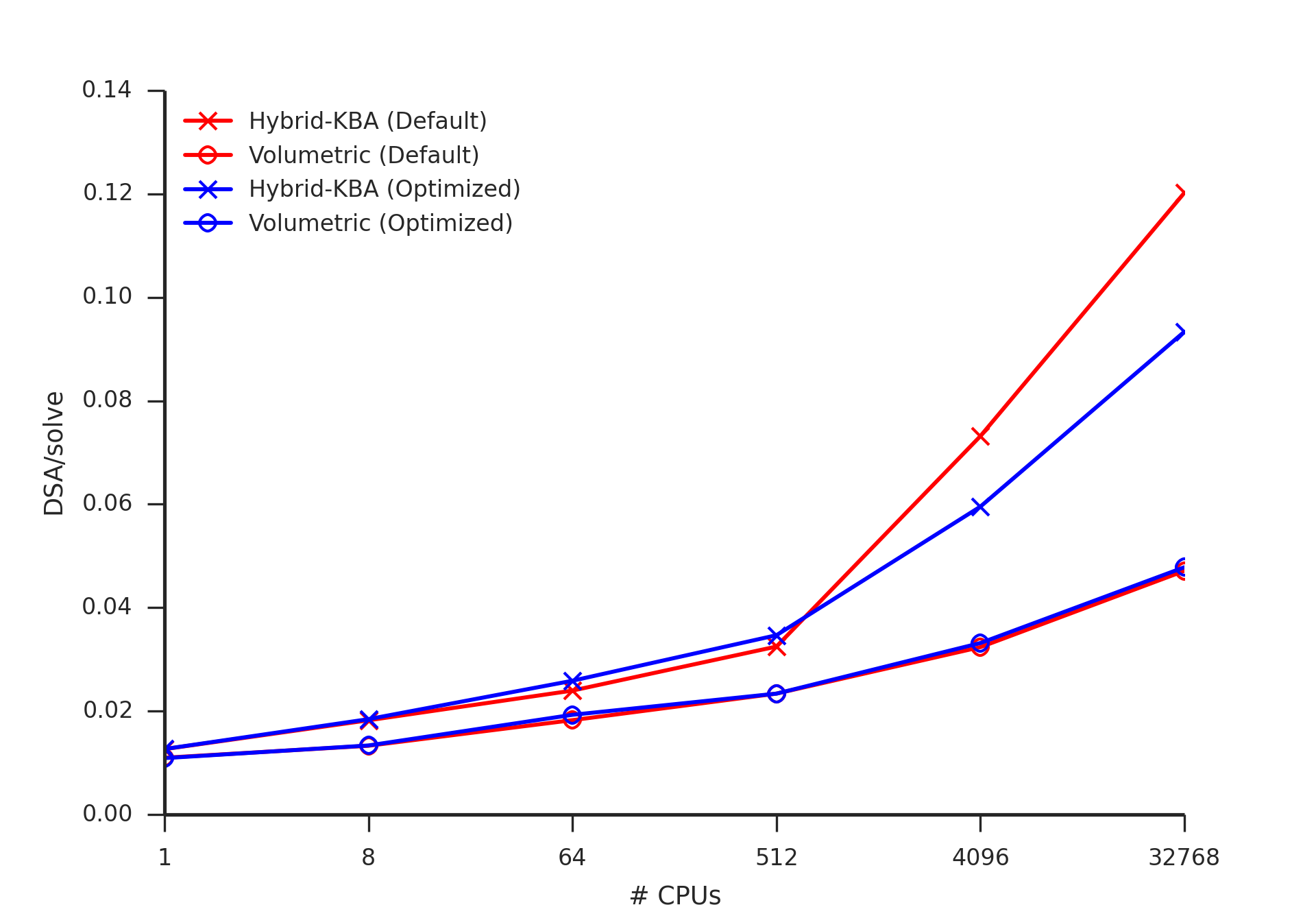}
\caption[]{\begin{tabular}[t]{l}%
	Hybrid-KBA vs. volumetric\\%
    DSA/solve fractions, 512 cells/core, resid\_thr = $10^{-4}$.%
\end{tabular}}
\label{fig:dsa2solve-512_vol}
\end{figure}

\begin{figure}[htb]
\centering
\includegraphics[scale=.7]{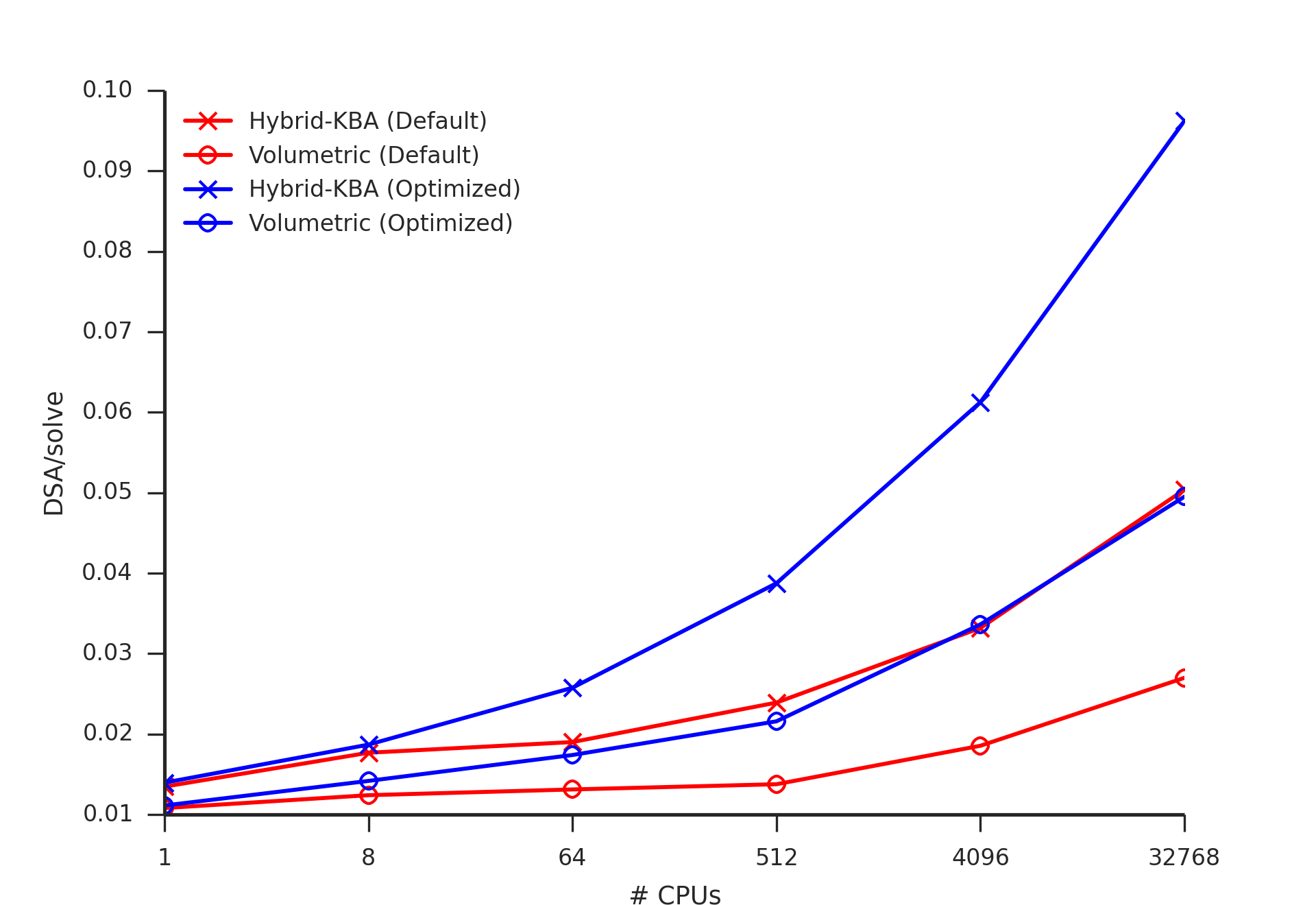}
\caption[]{\begin{tabular}[t]{l}%
	Hybrid-KBA vs. volumetric\\%
    DSA/solve fractions, 4096 cells/core, resid\_thr = $10^{-4}$.%
\end{tabular}}

\label{fig:dsa2solve-4096_vol}
\end{figure}

As expected, the DSA performed significantly better in this configuration, but the transport sweep times suffered and essentially swamped any gains from the faster DSA times (see Fig. \ref{fig:default-512_2048-vs-default-512_2048_vol} and \ref{fig:default-4096_2048-vs-default-4096_2048_vol}). For the optimized settings, we only show graphs of just the DSA times, better showing the effect of changing parallel partitioning  on HYPRE performance (Fig. \ref{fig:opt_resid-512_2048-vs-opt_resid-512_2048_vol} and \ref{fig:opt_resid-4096_2048-vs-opt_resid-4096_2048_vol_hypre}). The DSA/total solve time ratios are compared in figures \ref{fig:dsa2solve-512_vol} and \ref{fig:dsa2solve-4096_vol}.

.

\clearpage


\section{Acknowledgements}
This material is based upon work supported by the Department of Energy, National Nuclear
Security Administration, under Award Number(s) DE-NA0002376. Established by Congress in 2000, NNSA is a semi-autonomous agency within the U.S. Department of Energy responsible for enhancing national security through the military application of nuclear science. NNSA maintains and enhances the safety, security, reliability and performance of the U.S. nuclear weapons stockpile without nuclear testing; works to reduce global danger from weapons of mass destruction; provides the U.S. Navy with safe and effective nuclear propulsion; and responds to nuclear and radiological emergencies in the U.S. and abroad.

\bibliographystyle{plain}
\bibliography{ref.bib}


\clearpage

\appendix
\section{Raw Data}

This appendix contains the raw data tables with the times, numbers of iterations and residuals from the performed runs. The times are reported in milliseconds. Other notes (see also the end of Sec. \ref{sec:def-param}):

\begin{itemize}
\item \textit{Overall} = \textit{Sweep Setup} + \textit{Sweep} + \textit{WG DSA} + unreported problem setup times
\item \textit{WG DSA} = \textit{PCG Setup} + \textit{PCG Solve} + unreported other DSA times (mat/vec assembly, etc.)
\item \textit{Resid} -- the transport residual attained after 10 source iterations
\item \textit{PCG Resid} -- residual of the PCG solver attained after the total of \textit{PCG it} iterations
\end{itemize}

\begin{sidewaystable}
\centering
\begin{tabular}{c|c|rcrrrrrrrr}
 & \#cores &\ch{Overall} & \ch{Sweep Setup} & \ch{Sweep} & \ch{WG DSA} & \ch{PCG Setup} & \ch{PCG Solve} & \ch{PCG it} & \ch{Resid} & \ch{PCG Resid} & \ch{DSA/Solve}
\\\hline
\rh{Default (max\_it)}
 & 1       &  12500.0    &    802.0     &    7450.0    &    2860.0    &     82.3     &    2370.0    &    100     & 1.04903e-06  & 2.36344e-14  &   0.253097  \\
 & 8       &  14360.0    &    830.9     &    7823.0    &    3636.0    &    175.4     &    2902.0    &    100     & 5.34904e-06  & 1.66507e-10  &   0.288801  \\
 & 64      &  17530.0    &    826.8     &    9006.0    &    5275.0    &    320.1     &    4293.0    &    100     & 0.000510479  & 7.96546e-09  &   0.340323  \\
 & 512     &  20670.0    &    865.8     &    9943.0    &    7157.0    &    782.0     &    5615.0    &    100     &  0.00739123  & 6.53788e-09  &   0.387913  \\
 & 1024    &  26170.0    &    897.0     &   10440.0    &   11810.0    &    1767.0    &    9274.0    &    100     &  0.0111011   & 5.32665e-11  &   0.499155  \\
 & 2048    &  27280.0    &    903.1     &   10960.0    &   11950.0    &    2150.0    &    8980.0    &    100     &  0.0112932   & 7.13979e-10  &   0.491163  \\
 & 4096    &  29220.0    &    884.2     &   11620.0    &   12460.0    &    2842.0    &    8764.0    &    100     &   0.011109   & 1.69716e-08  &   0.488053  \\
 & 8192    &  37670.0    &    959.9     &   12260.0    &   19810.0    &    4889.0    &   14040.0    &    100     &  0.0111101   & 2.56958e-09  &   0.588882  \\
 & 16384   &  41550.0    &    965.8     &   13550.0    &   22340.0    &    6741.0    &   14700.0    &    100     &   0.010726   & 3.94823e-09  &   0.596369  \\
 & 32768   &  45850.0    &    958.0     &   14270.0    &   25440.0    &    8642.0    &   15860.0    &    100     &  0.00914706  & 3.92257e-07  &   0.615981  \\
 & 65536   &  57760.0    &    966.9     &   15770.0    &   35300.0    &   12520.0    &   21850.0    &    100     &  0.00914748  & 7.32815e-08  &   0.670084  \\
 & 131072  &  70480.0    &    1115.0    &   17490.0    &   44440.0    &   17430.0    &   26050.0    &    100     &  0.00891176  & 3.02956e-07  &   0.697098  \\
 \hline
\rh{Default (resid)}
 & 1       &  11500.0    &    809.0     &    7430.0    &    1570.0    &     82.2     &    1080.0    &     40     & 1.04922e-06  & 1.08824e-05  &    0.157    \\
 & 8       &  13660.0    &    837.8     &    7873.0    &    2320.0    &    176.5     &    1585.0    &     50     & 5.34917e-06  & 2.29012e-05  &   0.204586  \\
 & 64      &  16160.0    &    827.2     &    9032.0    &    3333.0    &    336.4     &    2339.0    &     50     & 0.000510497  & 8.25435e-05  &   0.245434  \\
 & 512     &  18300.0    &    857.1     &    9970.0    &    4613.0    &    787.4     &    3081.0    &     50     &  0.00739124  & 4.73157e-05  &   0.289761  \\
 & 1024    &  22680.0    &    889.8     &   10430.0    &    7097.0    &    1762.0    &    4577.0    &     44     &  0.0111011   & 1.59773e-05  &   0.374908  \\
 & 2048    &  22620.0    &    900.3     &   10940.0    &    7395.0    &    2123.0    &    4469.0    &     45     &  0.0112932   & 3.67704e-05  &   0.37443   \\
 & 4096    &  26020.0    &    881.4     &   11630.0    &    9201.0    &    2849.0    &    5518.0    &     58     &   0.011109   & 2.28592e-05  &   0.413157  \\
 & 8192    &  33460.0    &    956.0     &   12250.0    &   13470.0    &    4898.0    &    7704.0    &     50     &  0.0111101   & 7.01449e-05  &   0.493949  \\
 & 16384   &  35710.0    &    955.4     &   13520.0    &   16550.0    &    6752.0    &    8899.0    &     56     &  0.0107261   &  3.093e-05   &   0.523237  \\
 & 32768   &  41420.0    &    958.8     &   14220.0    &   21160.0    &    8697.0    &   11550.0    &     69     &  0.00914707  & 4.92867e-05  &   0.572201  \\
 & 65536   &  51370.0    &    951.4     &   15710.0    &   28410.0    &   12500.0    &   14980.0    &     65     &  0.00914745  & 9.52665e-05  &   0.621119  \\
 & 131072  &  64530.0    &    1115.0    &   17470.0    &   37840.0    &   17500.0    &   19390.0    &     71     &  0.00891174  & 8.73188e-05  &   0.662349  \\
 \hline
\rh{Optimized (resid)}
 & 1       &  11200.0    &    817.0     &    7390.0    &    1560.0    &     69.4     &    1090.0    &     41     & 1.04903e-06  & 1.13045e-05  &   0.156627  \\
 & 8       &  13010.0    &    839.2     &    7770.0    &    2338.0    &    137.5     &    1645.0    &     54     & 5.34948e-06  & 2.61807e-05  &   0.207638  \\
 & 64      &  15790.0    &    827.3     &    9036.0    &    3491.0    &    241.9     &    2589.0    &     66     &  0.00051053  & 3.84888e-05  &   0.253891  \\
 & 512     &  18460.0    &    859.8     &    9957.0    &    4921.0    &    357.6     &    3808.0    &     90     &  0.00739134  & 5.16197e-05  &   0.303204  \\
 & 1024    &  21710.0    &    893.2     &   10440.0    &    7492.0    &    577.9     &    6166.0    &    112     &  0.0111013   & 6.69757e-05  &   0.387384  \\
 & 2048    &  23020.0    &    892.6     &   10920.0    &    7948.0    &    586.4     &    6545.0    &    124     &  0.0112932   & 6.39623e-05  &   0.391527  \\
 & 4096    &  24260.0    &    881.3     &   11620.0    &    8250.0    &    537.6     &    6865.0    &    147     &  0.0111092   &  9.9524e-05  &   0.386961  \\
 & 8192    &  31500.0    &    949.0     &   12210.0    &   13790.0    &    916.2     &   12000.0    &    186     &  0.0111102   & 8.64342e-05  &     0.5     \\
 & 16384   &  34380.0    &    961.0     &   13510.0    &   14590.0    &    916.0     &   12760.0    &    207     &  0.0107261   & 7.33212e-05  &   0.491577  \\
 & 32768   &  35450.0    &    967.1     &   14200.0    &   15200.0    &    896.2     &   13380.0    &    241     &  0.00914717  & 8.24186e-05  &   0.490164  \\
 & 65536   &  51550.0    &    959.4     &   15640.0    &   29070.0    &    1519.0    &   26610.0    &    307     &  0.00914752  & 8.36885e-05  &   0.627049  \\
 & 131072  &  57420.0    &    1124.0    &   17380.0    &   31330.0    &    1605.0    &   28750.0    &    334     &  0.00891182  & 8.39437e-05  &   0.61966   \\
\end{tabular}
\caption{512 cells/core, 128 directions, hybrid-KBA}
\end{sidewaystable}

\begin{sidewaystable}
\centering
\begin{tabular}{c|c|rcrrrrrrrr}
 & \#cores &\ch{Overall} & \ch{Sweep Setup} & \ch{Sweep} & \ch{WG DSA} & \ch{PCG Setup} & \ch{PCG Solve} & \ch{PCG it} & \ch{Resid} & \ch{PCG Resid} & \ch{DSA/Solve}
\\\hline
\rh{Default (max\_it)} 
 & 1       &  35500.0    &    3130.0    &   27800.0    &    2860.0    &     82.3     &    2370.0    &    100     & 1.07371e-06  & 2.37575e-14  &  0.0841176  \\
 & 8       &  37820.0    &    3164.0    &   28600.0    &    3637.0    &    176.7     &    2904.0    &    100     & 5.54237e-06  & 1.65335e-10  &   0.101877  \\
 & 64      &  42600.0    &    3201.0    &   31580.0    &    5052.0    &    324.3     &    4058.0    &    100     &  0.00054167  & 1.04996e-09  &   0.125578  \\
 & 512     &  47260.0    &    3196.0    &   33860.0    &    7165.0    &    780.8     &    5619.0    &    100     &  0.00758877  & 6.43563e-09  &   0.16004   \\
 & 1024    &  56310.0    &    3263.0    &   33830.0    &   15910.0    &    1907.0    &   13230.0    &    100     &  0.0115247   & 5.24375e-11  &   0.297272  \\
 & 2048    &  53770.0    &    3243.0    &   35400.0    &   11820.0    &    2118.0    &    8880.0    &    100     &   0.011589   & 6.99219e-10  &   0.231856  \\
 & 4096    &  61500.0    &    3296.0    &   37650.0    &   15980.0    &    3929.0    &   11200.0    &    100     &  0.0113932   & 8.14588e-09  &   0.278058  \\
 & 8192    &  67630.0    &    3323.0    &   37430.0    &   22070.0    &    5080.0    &   16120.0    &    100     &  0.0113943   & 2.53077e-09  &   0.347997  \\
 & 16384   &  70770.0    &    3315.0    &   39820.0    &   22560.0    &    6730.0    &   14900.0    &    100     &  0.0109988   &  3.9089e-09  &   0.340169  \\
 & 32768   &  76030.0    &    3317.0    &   41990.0    &   25450.0    &    8682.0    &   15820.0    &    100     &  0.00943057  & 3.81094e-07  &   0.356443  \\
 & 65536   &  92270.0    &    3470.0    &   45680.0    &   36600.0    &   12470.0    &   23180.0    &    100     &  0.00943101  & 1.27278e-07  &   0.423317  \\
 & 131072  &  104100.0   &    3468.0    &   47800.0    &   45410.0    &   17470.0    &   26980.0    &    100     &  0.00917903  & 2.96668e-07  &   0.465935  \\
 \hline
 \rh{Default (resid)}
 & 1       &  34400.0    &    3140.0    &   28000.0    &    1570.0    &     82.4     &    1080.0    &     40     &  1.0739e-06  & 1.08638e-05  &  0.0477204  \\
 & 8       &  36570.0    &    3161.0    &   28760.0    &    2315.0    &    174.2     &    1585.0    &     50     & 5.54251e-06  & 2.29668e-05  &  0.0670237  \\
 & 64      &  40920.0    &    3167.0    &   31760.0    &    3294.0    &    325.5     &    2302.0    &     52     & 0.000541706  & 1.70803e-05  &  0.0852926  \\
 & 512     &  44730.0    &    3193.0    &   34040.0    &    4643.0    &    789.9     &    3100.0    &     50     &  0.00758879  & 4.65253e-05  &   0.109608  \\
 & 1024    &  49580.0    &    3249.0    &   34010.0    &    9127.0    &    1901.0    &    6458.0    &     44     &  0.0115248   & 1.57449e-05  &   0.194606  \\
 & 2048    &  49730.0    &    3226.0    &   35410.0    &    7469.0    &    2138.0    &    4513.0    &     45     &  0.0115889   & 3.60819e-05  &   0.160142  \\
 & 4096    &  57610.0    &    3269.0    &   37740.0    &   11630.0    &    3950.0    &    6829.0    &     57     &  0.0113933   & 4.12366e-05  &   0.218527  \\
 & 8192    &  62210.0    &    3296.0    &   37570.0    &   13970.0    &    4961.0    &    8129.0    &     50     &  0.0113943   &  6.9492e-05  &   0.251984  \\
 & 16384   &  64790.0    &    3292.0    &   39970.0    &   16610.0    &    6758.0    &    8926.0    &     56     &  0.0109988   & 3.03171e-05  &   0.274545  \\
 & 32768   &  71970.0    &    3289.0    &   42100.0    &   21200.0    &    8683.0    &   11580.0    &     69     &  0.00943058  & 4.81038e-05  &   0.315289  \\
 & 65536   &  85290.0    &    3457.0    &   45400.0    &   29770.0    &   12440.0    &   16380.0    &     68     &  0.00943097  & 3.34576e-05  &   0.375268  \\
 & 131072  &  98640.0    &    3466.0    &   47990.0    &   39410.0    &   17660.0    &   20790.0    &     71     &  0.00917902  & 8.55293e-05  &   0.430005  \\
 \hline
\rh{Optimized (resid)}
 & 1       &  34100.0    &    3150.0    &   27800.0    &    1560.0    &     69.3     &    1090.0    &     41     & 1.07371e-06  & 1.12791e-05  &  0.0478528  \\
 & 8       &  36100.0    &    3153.0    &   28330.0    &    2315.0    &    136.7     &    1619.0    &     53     & 5.54271e-06  & 2.63684e-05  &  0.0678687  \\
 & 64      &  40890.0    &    3164.0    &   31510.0    &    3532.0    &    237.5     &    2624.0    &     68     & 0.000541674  & 2.75909e-05  &  0.0914789  \\
 & 512     &  44630.0    &    3193.0    &   33630.0    &    4918.0    &    360.1     &    3799.0    &     90     &  0.00758889  & 5.11703e-05  &   0.116457  \\
 & 1024    &  48230.0    &    3250.0    &   33920.0    &    7937.0    &    626.0     &    6557.0    &    112     &   0.011525   & 6.58878e-05  &   0.174019  \\
 & 2048    &  50040.0    &    3226.0    &   35110.0    &    8180.0    &    595.4     &    6757.0    &    124     &   0.011589   & 6.27538e-05  &   0.173784  \\
 & 4096    &  53710.0    &    3268.0    &   37460.0    &    9119.0    &    662.4     &    7589.0    &    151     &  0.0113937   & 7.17866e-05  &   0.180825  \\
 & 8192    &  62380.0    &    3289.0    &   37290.0    &   17040.0    &    1155.0    &   15000.0    &    186     &  0.0113944   & 8.70697e-05  &   0.292482  \\
 & 16384   &  61740.0    &    3296.0    &   39610.0    &   13940.0    &    859.3     &   12150.0    &    207     &  0.0109989   &  7.1832e-05  &   0.242435  \\
 & 32768   &  65950.0    &    3296.0    &   41830.0    &   15540.0    &    907.3     &   13690.0    &    241     &  0.00943068  & 8.01564e-05  &   0.253425  \\
 & 65536   &  84330.0    &    3469.0    &   45140.0    &   29020.0    &    1879.0    &   26190.0    &    315     &  0.00943104  & 7.57143e-05  &   0.370389  \\
 & 131072  &  90480.0    &    3474.0    &   47450.0    &   31820.0    &    1608.0    &   29240.0    &    334     &  0.0091791   & 8.21673e-05  &   0.380941  \\
\end{tabular}
\caption{512 cells/core, 512 directions, hybrid-KBA}
\end{sidewaystable}

\begin{sidewaystable}
\centering
\begin{tabular}{c|c|rcrrrrrrrr}
 & \#cores &\ch{Overall} & \ch{Sweep Setup} & \ch{Sweep} & \ch{WG DSA} & \ch{PCG Setup} & \ch{PCG Solve} & \ch{PCG it} & \ch{Resid} & \ch{PCG Resid} & \ch{DSA/Solve}
\\\hline
\rh{Default} 
 & 1       &  128000.0   &   12200.0    &   110000.0   &    2870.0    &     82.2     &    2370.0    &    100     & 1.07974e-06  & 2.37906e-14  &  0.0227778  \\
 & 8       &  131300.0   &   12250.0    &   112000.0   &    3649.0    &    174.4     &    2906.0    &    100     & 5.58992e-06  & 1.65124e-10  &  0.0284633  \\
 & 64      &  142100.0   &   12320.0    &   121000.0   &    5062.0    &    323.6     &    4066.0    &    100     & 0.000549206  & 1.04522e-09  &  0.0364697  \\
 & 512     &  148600.0   &   12380.0    &   125000.0   &    7252.0    &    785.0     &    5710.0    &    100     &  0.00763565  & 6.41056e-09  &  0.0499793  \\
 & 1024    &  156900.0   &   12460.0    &   124800.0   &   15410.0    &    1914.0    &   12720.0    &    100     &  0.0116384   & 5.22424e-11  &   0.100587  \\
 & 2048    &  157000.0   &   12400.0    &   128200.0   &   12120.0    &    2133.0    &    9160.0    &    100     &  0.0116591   & 6.95476e-10  &  0.0790607  \\
 & 4096    &  167300.0   &   12500.0    &   133300.0   &   15960.0    &    3938.0    &   11170.0    &    100     &  0.0114607   & 8.10848e-09  &  0.0983364  \\
 & 8192    &  176000.0   &   12510.0    &   133500.0   &   24120.0    &    5053.0    &   18180.0    &    100     &  0.0114618   & 2.52214e-09  &   0.141218  \\
 & 16384   &  179200.0   &   12520.0    &   137800.0   &   22830.0    &    6751.0    &   15150.0    &    100     &  0.0110636   & 3.90361e-09  &   0.131358  \\
 & 32768   &  186700.0   &   12500.0    &   142000.0   &   25940.0    &    8690.0    &   16300.0    &    100     &  0.00949775  & 3.78613e-07  &   0.143236  \\
 & 65536   &  219800.0   &   12740.0    &   149000.0   &   50430.0    &   12860.0    &   36600.0    &    100     &  0.00949819  & 1.26525e-07  &   0.236761  \\
 & 131072  &  219700.0   &   12710.0    &   153300.0   &   45430.0    &   17480.0    &   26950.0    &    100     &  0.00924395  & 2.95089e-07  &   0.21399   \\
 \hline
\rh{Optimized} 
 & 1       &  128000.0   &   12200.0    &   111000.0   &    2830.0    &     69.3     &    2350.0    &    100     & 1.07974e-06  & 1.73123e-13  &  0.0224603  \\
 & 8       &  130800.0   &   12250.0    &   111700.0   &    3533.0    &    137.1     &    2826.0    &    100     & 5.58992e-06  &  6.0878e-10  &  0.0276448  \\
 & 64      &  141300.0   &   12310.0    &   120700.0   &    4586.0    &    236.8     &    3680.0    &    100     &  0.0005492   & 2.24944e-07  &  0.0332319  \\
 & 512     &  146900.0   &   12380.0    &   125300.0   &    5338.0    &    358.7     &    4220.0    &    100     &  0.00763821  &  5.4886e-06  &  0.0372245  \\
 & 1024    &  150600.0   &   12420.0    &   125100.0   &    8901.0    &    791.5     &    7340.0    &    100     &  0.0116623   & 3.31873e-05  &  0.0605922  \\
 & 2048    &  151800.0   &   12400.0    &   128000.0   &    6897.0    &    591.4     &    5474.0    &    100     &  0.0117057   & 6.25597e-05  &  0.0466329  \\
 & 4096    &  157100.0   &   12490.0    &   133000.0   &    6765.0    &    658.2     &    5238.0    &    100     &   0.011534   & 0.000184994  &  0.0442736  \\
 & 8192    &  162500.0   &   12490.0    &   133400.0   &   10910.0    &    1179.0    &    8835.0    &    100     &   0.011464   & 0.000392088  &  0.0693139  \\
 & 16384   &  165100.0   &   12520.0    &   137700.0   &    8512.0    &    931.4     &    6644.0    &    100     &  0.0114799   &  0.00082181  &  0.0534003  \\
 & 32768   &  168900.0   &   12500.0    &   142100.0   &    8048.0    &    918.3     &    6180.0    &    100     &  0.00999497  &  0.00374776  &  0.0492534  \\
 & 65536   &  185100.0   &   12730.0    &   149000.0   &   15870.0    &    2216.0    &   12680.0    &    100     &  0.00971854  &  0.0673185   &  0.0889574  \\
 & 131072  &  186900.0   &   12710.0    &   153600.0   &   12240.0    &    1653.0    &    9583.0    &    100     &  0.0104379   &   0.114973   &  0.0682655  \\
\end{tabular}
\caption{512 cells/core, 2048 directions, hybrid-KBA, PCG max\_it = 10}
\end{sidewaystable}

\begin{sidewaystable}
\centering
\begin{tabular}{c|c|rcrrrrrrrr}
 & \#cores &\ch{Overall} & \ch{Sweep Setup} & \ch{Sweep} & \ch{WG DSA} & \ch{PCG Setup} & \ch{PCG Solve} & \ch{PCG it} & \ch{Resid} & \ch{PCG Resid} & \ch{DSA/Solve}
\\\hline
\rh{Default} 
 & 1       &  128000.0   &   12200.0    &   111000.0   &    1580.0    &     82.2     &    1080.0    &     40     & 1.07993e-06  &  1.0859e-05  &   0.01264   \\
 & 8       &  130700.0   &   12250.0    &   112800.0   &    2327.0    &    174.2     &    1584.0    &     50     & 5.59006e-06  &  2.2993e-05  &  0.0182224  \\
 & 64      &  140900.0   &   12310.0    &   121600.0   &    3297.0    &    325.2     &    2303.0    &     52     & 0.000549243  & 1.70031e-05  &  0.0239608  \\
 & 512     &  146900.0   &   12360.0    &   126000.0   &    4660.0    &    787.1     &    3121.0    &     50     &  0.00763566  & 4.63282e-05  &  0.0324739  \\
 & 1024    &  151200.0   &   12410.0    &   125600.0   &    8940.0    &    1914.0    &    6259.0    &     44     &  0.0116384   & 1.56899e-05  &  0.0606102  \\
 & 2048    &  153000.0   &   12390.0    &   128500.0   &    7513.0    &    2161.0    &    4538.0    &     45     &  0.0116591   & 3.58998e-05  &  0.0504228  \\
 & 4096    &  163600.0   &   12480.0    &   133700.0   &   11590.0    &    3939.0    &    6795.0    &     57     &  0.0114608   & 4.10403e-05  &  0.0731692  \\
 & 8192    &  168000.0   &   12480.0    &   133800.0   &   15970.0    &    5087.0    &    9995.0    &     50     &  0.0114618   & 6.93691e-05  &  0.0980356  \\
 & 16384   &  173800.0   &   12500.0    &   138600.0   &   16800.0    &    6755.0    &    9127.0    &     56     &  0.0110636   & 3.01657e-05  &  0.0996441  \\
 & 32768   &  182600.0   &   12490.0    &   142500.0   &   21280.0    &    8680.0    &   11650.0    &     69     &  0.00949776  & 4.78138e-05  &   0.120226  \\
 & 65536   &  202900.0   &   12710.0    &   149400.0   &   32820.0    &   12550.0    &   19300.0    &     68     &  0.00949815  & 3.32758e-05  &   0.16762   \\
 & 131072  &  214700.0   &   12690.0    &   154400.0   &   38730.0    &   17660.0    &   20080.0    &     71     &  0.00924394  & 8.50846e-05  &   0.187464  \\
 \hline
\rh{Optimized} 
 & 1       &  126000.0   &   12200.0    &   110000.0   &    1570.0    &     69.4     &    1090.0    &     41     & 1.07974e-06  & 1.12727e-05  &  0.0126613  \\
 & 8       &  128800.0   &   12240.0    &   111000.0   &    2321.0    &    137.2     &    1620.0    &     53     & 5.59029e-06  & 2.65561e-05  &  0.0184353  \\
 & 64      &  139700.0   &   12290.0    &   120100.0   &    3528.0    &    236.4     &    2622.0    &     68     & 0.000549211  & 2.75039e-05  &  0.0258651  \\
 & 512     &  145800.0   &   12370.0    &   124700.0   &    4940.0    &    356.0     &    3824.0    &     90     &  0.00763576  & 5.10912e-05  &  0.0346667  \\
 & 1024    &  150600.0   &   12410.0    &   124500.0   &    9629.0    &    798.3     &    8069.0    &    112     &  0.0116387   & 6.56522e-05  &  0.0655034  \\
 & 2048    &  152500.0   &   12400.0    &   127300.0   &    8250.0    &    615.5     &    6804.0    &    124     &  0.0116592   & 6.24734e-05  &  0.0555556  \\
 & 4096    &  158900.0   &   12490.0    &   132400.0   &    9205.0    &    663.6     &    7668.0    &    151     &  0.0114612   &  7.1473e-05  &  0.0595408  \\
 & 8192    &  169100.0   &   12490.0    &   132600.0   &   18270.0    &    1280.0    &   16100.0    &    186     &  0.0114619   & 8.73392e-05  &   0.111402  \\
 & 16384   &  169800.0   &   12520.0    &   137000.0   &   14310.0    &    882.1     &   12490.0    &    207     &  0.0110637   & 7.14696e-05  &  0.0869909  \\
 & 32768   &  175900.0   &   12500.0    &   141100.0   &   15890.0    &    921.8     &   14010.0    &    241     &  0.00949786  & 7.95755e-05  &  0.0933608  \\
 & 65536   &  211700.0   &   12730.0    &   149100.0   &   42000.0    &    2312.0    &   38690.0    &    315     &  0.00949822  & 7.58443e-05  &   0.205178  \\
 & 131072  &  207000.0   &   12710.0    &   152700.0   &   32740.0    &    1649.0    &   30090.0    &    334     &  0.00924402  & 8.17431e-05  &   0.164605  \\
\end{tabular}
\caption{512 cells/core, 2048 directions, hybrid-KBA, PCG resid\_thr = $10^{-4}$}
\end{sidewaystable}

\begin{sidewaystable}
\centering
\begin{tabular}{c|c|rcrrrrrrrr}
 & \#cores &\ch{Overall} & \ch{Sweep Setup} & \ch{Sweep} & \ch{WG DSA} & \ch{PCG Setup} & \ch{PCG Solve} & \ch{PCG it} & \ch{Resid} & \ch{PCG Resid} & \ch{DSA/Solve}
\\\hline
\rh{Default (max\_it} 
 & 1       &  101000.0   &    6150.0    &   60900.0    &   24100.0    &    801.0     &   19900.0    &    100     & 5.34904e-06  & 2.89473e-13  &   0.259978  \\
 & 8       &  128300.0   &    6232.0    &   63370.0    &   30850.0    &    1224.0    &   25870.0    &    100     & 0.000510479  & 1.85541e-09  &   0.301859  \\
 & 64      &  122300.0   &    6287.0    &   67930.0    &   34360.0    &    1626.0    &   28790.0    &    100     &  0.00739123  & 4.01045e-09  &   0.310669  \\
 & 512     &  129300.0   &    6332.0    &   71000.0    &   37060.0    &    2277.0    &   30730.0    &    100     &   0.011109   & 1.13482e-08  &   0.318385  \\
 & 1024    &  162900.0   &    6430.0    &   71580.0    &   69620.0    &    5650.0    &   59920.0    &    100     &  0.0111101   & 6.29845e-09  &   0.464753  \\
 & 2048    &  145200.0   &    6439.0    &   72920.0    &   49830.0    &    4001.0    &   41660.0    &    100     &   0.010726   & 2.74078e-09  &   0.379513  \\
 & 4096    &  142800.0   &    6380.0    &   75220.0    &   44390.0    &    4842.0    &   35340.0    &    100     &  0.00914706  & 1.50376e-07  &   0.346256  \\
 & 8192    &  185600.0   &    6593.0    &   77800.0    &   82620.0    &   10870.0    &   67490.0    &    100     &  0.00914749  & 1.85122e-07  &   0.487721  \\
 & 16384   &  168100.0   &    6728.0    &   79370.0    &   62630.0    &    9681.0    &   48660.0    &    100     &  0.00891176  & 1.92608e-07  &   0.414768  \\
 & 32768   &  171100.0   &    6691.0    &   81990.0    &   60820.0    &   12430.0    &   44070.0    &    100     &  0.00787385  & 1.21658e-05  &   0.400659  \\
 & 65536   &  232100.0   &    6977.0    &   88940.0    &   108800.0   &   26800.0    &   77620.0    &    100     &  0.00787421  & 7.42293e-06  &   0.524843  \\
 & 131072  &  229200.0   &    7062.0    &   91710.0    &   96550.0    &   23810.0    &   68290.0    &    100     &  0.00780639  & 5.61631e-06  &   0.488119  \\
 \hline
\rh{Default (resid)}
 & 1       &  91400.0    &    6270.0    &   61400.0    &   13500.0    &    795.0     &    9390.0    &     42     &  5.3501e-06  & 1.52647e-05  &   0.163438  \\
 & 8       &  116300.0   &    6338.0    &   63540.0    &   18050.0    &    1204.0    &   13160.0    &     46     & 0.000510495  & 3.21374e-05  &   0.201316  \\
 & 64      &  108400.0   &    6344.0    &   68240.0    &   20430.0    &    1633.0    &   14940.0    &     47     &  0.00739118  & 2.28238e-05  &   0.210901  \\
 & 512     &  115000.0   &    6333.0    &   71040.0    &   23000.0    &    2276.0    &   16770.0    &     50     &   0.011109   & 8.81729e-05  &   0.224609  \\
 & 1024    &  144600.0   &    6485.0    &   71920.0    &   43590.0    &    6391.0    &   33220.0    &     51     &  0.0111101   &  6.8356e-05  &   0.351249  \\
 & 2048    &  128000.0   &    6427.0    &   73260.0    &   32780.0    &    3956.0    &   24750.0    &     56     &  0.0107261   & 2.35431e-05  &   0.286288  \\
 & 4096    &  134000.0   &    6387.0    &   75400.0    &   34540.0    &    4889.0    &   25550.0    &     69     &  0.00914706  & 4.49515e-05  &   0.291477  \\
 & 8192    &  168300.0   &    6581.0    &   77830.0    &   63020.0    &   10880.0    &   47970.0    &     68     &  0.00914745  & 4.11459e-05  &   0.420975  \\
 & 16384   &  158300.0   &    6719.0    &   80100.0    &   48920.0    &    9717.0    &   35000.0    &     69     &  0.00891175  & 6.30107e-05  &   0.354493  \\
 & 32768   &  171100.0   &    6678.0    &   82210.0    &   59970.0    &   12600.0    &   43140.0    &     97     &  0.00787385  & 3.75653e-05  &   0.396627  \\
 & 65536   &  226900.0   &    6961.0    &   89050.0    &   101800.0   &   26770.0    &   70750.0    &     90     &  0.00787413  &  6.1537e-05  &   0.508238  \\
 & 131072  &  228300.0   &    7066.0    &   92190.0    &   90640.0    &   23880.0    &   62360.0    &     90     &  0.0078064   & 6.67985e-05  &   0.471347  \\
 \hline
\rh{Optimized (resid)} 
 & 1       &  91400.0    &    6380.0    &   60900.0    &   13900.0    &    632.0     &    9900.0    &     46     & 5.34946e-06  & 1.55027e-05  &   0.168077  \\
 & 8       &  117900.0   &    6375.0    &   63080.0    &   20740.0    &    900.4     &   16150.0    &     60     & 0.000510771  & 8.10579e-05  &   0.225557  \\
 & 64      &  114900.0   &    6344.0    &   68130.0    &   27180.0    &    1171.0    &   22140.0    &     77     &  0.00739133  & 7.39023e-05  &   0.262355  \\
 & 512     &  131000.0   &    6336.0    &   71210.0    &   38890.0    &    1363.0    &   33530.0    &    118     &  0.0111093   & 7.51212e-05  &   0.328186  \\
 & 1024    &  169500.0   &    6510.0    &   71480.0    &   76250.0    &    2532.0    &   69750.0    &    159     &  0.0111104   & 8.98174e-05  &   0.487532  \\
 & 2048    &  167300.0   &    6435.0    &   72860.0    &   72360.0    &    2449.0    &   65800.0    &    181     &   0.010726   & 7.04623e-05  &   0.470481  \\
 & 4096    &  164200.0   &    6396.0    &   75000.0    &   66270.0    &    1722.0    &   60410.0    &    210     &  0.00914711  & 8.95374e-05  &   0.442095  \\
 & 8192    &  235200.0   &    6604.0    &   77640.0    &   132400.0   &    3563.0    &   124700.0   &    275     &  0.00914754  & 7.18688e-05  &   0.60429   \\
 & 16384   &  220700.0   &    6741.0    &   79550.0    &   114900.0   &    2975.0    &   107700.0   &    301     &  0.00891179  & 7.28596e-05  &   0.564619  \\
 & 32768   &  224600.0   &    6711.0    &   81620.0    &   114100.0   &    2341.0    &   107500.0   &    354     &  0.00787387  & 9.08799e-05  &   0.557129  \\
 & 65536   &  349200.0   &    7035.0    &   88520.0    &   224600.0   &    9851.0    &   210400.0   &    454     &  0.00787413  & 8.95597e-05  &   0.696002  \\
 & 131072  &  373800.0   &    7083.0    &   91230.0    &   237600.0   &    4946.0    &   228300.0   &    496     &  0.00780644  & 9.98562e-05  &   0.702128  \\
\end{tabular}
\caption{4096 cells/core, 128 directions, hybrid-KBA}
\end{sidewaystable}

\begin{sidewaystable}
\centering
\begin{tabular}{c|c|rcrrrrrrrr}
& \#cores &\ch{Overall} & \ch{Sweep Setup} & \ch{Sweep} & \ch{WG DSA} & \ch{PCG Setup} & \ch{PCG Solve} & \ch{PCG it} & \ch{Resid} & \ch{PCG Resid} & \ch{DSA/Solve}
\\\hline
\rh{Default (max\_it)}
 & 1       &  287000.0   &   24800.0    &   226000.0   &   24000.0    &    795.0     &   19900.0    &    100     & 5.54237e-06  &  2.9486e-13  &  0.0869565  \\
 & 8       &  315300.0   &   24900.0    &   229800.0   &   30810.0    &    1223.0    &   25860.0    &    100     &  0.00054167  & 1.81016e-09  &   0.10724   \\
 & 64      &  319800.0   &   25070.0    &   244800.0   &   34650.0    &    1872.0    &   28860.0    &    100     &  0.00758877  & 1.88624e-10  &   0.113087  \\
 & 512     &  333600.0   &   25240.0    &   252600.0   &   38680.0    &    2700.0    &   31940.0    &    100     &  0.0113932   & 8.43999e-09  &   0.121406  \\
 & 1024    &  368200.0   &   25410.0    &   251800.0   &   73460.0    &    8119.0    &   61270.0    &    100     &  0.0113943   & 6.89169e-09  &   0.208279  \\
 & 2048    &  353700.0   &   25460.0    &   257800.0   &   52060.0    &    5076.0    &   42830.0    &    100     &  0.0109988   & 5.61461e-09  &   0.154298  \\
 & 4096    &  355700.0   &   25440.0    &   262400.0   &   49340.0    &    6347.0    &   38800.0    &    100     &  0.00943057  &  1.0959e-07  &   0.145417  \\
 & 8192    &  395400.0   &   25440.0    &   264000.0   &   85280.0    &   11050.0    &   70010.0    &    100     &  0.00943101  & 1.81165e-07  &   0.226147  \\
 & 16384   &  394700.0   &   25800.0    &   271000.0   &   76090.0    &   14320.0    &   57440.0    &    100     &  0.00917904  & 3.42193e-07  &   0.202799  \\
 & 32768   &  410000.0   &   25850.0    &   281500.0   &   77970.0    &   19520.0    &   54060.0    &    100     &  0.00812518  & 6.63397e-06  &   0.201109  \\
 & 65536   &  448400.0   &   25880.0    &   281600.0   &   111600.0   &   27160.0    &   80130.0    &    100     &  0.00812554  & 7.29787e-06  &   0.264706  \\
 & 131072  &  446800.0   &   25930.0    &   288600.0   &   96440.0    &   23750.0    &   68250.0    &    100     &  0.00804959  &  5.4844e-06  &   0.233285  \\
 \hline
\rh{Default (resid)}
 & 1       &  277000.0   &   24900.0    &   227000.0   &   13500.0    &    795.0     &    9390.0    &     42     & 5.54345e-06  & 1.52623e-05  &  0.0505618  \\
 & 8       &  304200.0   &   25080.0    &   231200.0   &   18070.0    &    1224.0    &   13160.0    &     46     & 0.000541686  & 3.19823e-05  &  0.0654473  \\
 & 64      &  306500.0   &   25010.0    &   245300.0   &   20400.0    &    1782.0    &   14730.0    &     46     &  0.00758876  &  2.6439e-05  &  0.0697198  \\
 & 512     &  321700.0   &   25170.0    &   253100.0   &   26430.0    &    2766.0    &   19640.0    &     57     &  0.0113934   & 3.92945e-05  &  0.0861193  \\
 & 1024    &  345300.0   &   25220.0    &   252600.0   &   47290.0    &    8180.0    &   35070.0    &     53     &  0.0113944   & 8.64778e-05  &   0.144485  \\
 & 2048    &  339400.0   &   25310.0    &   259700.0   &   35420.0    &    5097.0    &   26180.0    &     57     &  0.0109988   & 3.03615e-05  &   0.109761  \\
 & 4096    &  345600.0   &   25320.0    &   264100.0   &   37690.0    &    6370.0    &   27160.0    &     67     &  0.00943057  & 3.59235e-05  &   0.11442   \\
 & 8192    &  375000.0   &   25500.0    &   264900.0   &   64290.0    &   11070.0    &   49020.0    &     68     &  0.00943097  & 4.01691e-05  &   0.180084  \\
 & 16384   &  382600.0   &   25680.0    &   272600.0   &   60430.0    &   14320.0    &   41800.0    &     70     &  0.00917904  & 8.14918e-05  &   0.167303  \\
 & 32768   &  407500.0   &   25680.0    &   282200.0   &   73250.0    &   19550.0    &   49330.0    &     90     &  0.00812517  & 6.29388e-05  &   0.190904  \\
 & 65536   &  442100.0   &   25770.0    &   282500.0   &   102600.0   &   27130.0    &   71190.0    &     90     &  0.00812545  & 6.02556e-05  &   0.248066  \\
 & 131072  &  446300.0   &   25880.0    &   289800.0   &   90540.0    &   23880.0    &   62240.0    &     90     &  0.0080496   & 6.53035e-05  &   0.221586  \\
 \hline
\rh{Optimized (resid)} 
 & 1       &  275000.0   &   24900.0    &   224000.0   &   13900.0    &    632.0     &    9900.0    &     46     & 5.54284e-06  & 1.55307e-05  &  0.0524528  \\
 & 8       &  304100.0   &   24870.0    &   228900.0   &   20780.0    &    901.9     &   16180.0    &     60     &  0.00054197  & 7.97816e-05  &  0.0752081  \\
 & 64      &  310500.0   &   24900.0    &   242600.0   &   27620.0    &    1214.0    &   22490.0    &     78     &  0.00758879  & 7.53747e-05  &  0.0929653  \\
 & 512     &  335800.0   &   25080.0    &   251100.0   &   42590.0    &    1522.0    &   37010.0    &    128     &  0.0113934   & 7.65995e-05  &   0.13272   \\
 & 1024    &  380000.0   &   25340.0    &   251200.0   &   85910.0    &    4621.0    &   77260.0    &    167     &  0.0113945   & 7.20037e-05  &   0.235628  \\
 & 2048    &  373500.0   &   25240.0    &   256500.0   &   73660.0    &    2727.0    &   66780.0    &    184     &  0.0109988   & 8.42845e-05  &   0.205984  \\
 & 4096    &  375600.0   &   25300.0    &   260600.0   &   70980.0    &    1955.0    &   64850.0    &    222     &  0.00943066  & 8.82642e-05  &   0.197606  \\
 & 8192    &  447500.0   &   25330.0    &   262900.0   &   138800.0   &    3890.0    &   130700.0   &    273     &  0.00943109  & 9.82894e-05  &   0.323166  \\
 & 16384   &  442800.0   &   25660.0    &   269600.0   &   125600.0   &    3477.0    &   117800.0   &    314     &  0.00917908  &  7.3526e-05  &   0.296786  \\
 & 32768   &  466500.0   &   25690.0    &   279700.0   &   135500.0   &    3160.0    &   128000.0   &    379     &  0.0081252   & 8.80789e-05  &   0.305593  \\
 & 65536   &  581100.0   &   25780.0    &   281100.0   &   243100.0   &   10440.0    &   228300.0   &    454     &  0.00812545  & 8.83663e-05  &   0.43992   \\
 & 131072  &  588900.0   &   25890.0    &   286800.0   &   236300.0   &    4975.0    &   226900.0   &    492     &  0.00804962  & 9.74965e-05  &   0.428468  \\
\end{tabular}
\caption{4096 cells/core, 512 directions, hybrid-KBA}
\end{sidewaystable}

\begin{sidewaystable}
\centering
\begin{tabular}{c|c|rcrrrrrrrr}
 & \#cores &\ch{Overall} & \ch{Sweep Setup} & \ch{Sweep} & \ch{WG DSA} & \ch{PCG Setup} & \ch{PCG Solve} & \ch{PCG it} & \ch{Resid} & \ch{PCG Resid} & \ch{DSA/Solve}
\\\hline
\rh{Default}
 & 1       &  1.03e+06   &   97300.0    &   890000.0   &   24000.0    &    796.0     &   19900.0    &    100     & 5.58992e-06  & 2.96587e-13  &  0.0237624  \\
 & 8       & 1.118e+06   &   97490.0    &   953000.0   &   30820.0    &    1228.0    &   25870.0    &    100     & 0.000549206  & 1.79778e-09  &   0.028458  \\
 & 64      & 1.108e+06   &   97700.0    &   953400.0   &   34730.0    &    1878.0    &   28900.0    &    100     &  0.00763565  & 1.87927e-10  &   0.031921  \\
 & 512     & 1.131e+06   &   98210.0    &   969600.0   &   38720.0    &    2707.0    &   31930.0    &    100     &  0.0114607   & 8.39906e-09  &  0.0349143  \\
 & 1024    & 1.157e+06   &   98100.0    &   961300.0   &   72130.0    &    8118.0    &   59940.0    &    100     &  0.0114618   & 6.85756e-09  &  0.0636067  \\
 & 2048    &  1.15e+06   &   98110.0    &   977500.0   &   49290.0    &    3946.0    &   41210.0    &    100     &  0.0110636   & 2.66825e-09  &  0.0437356  \\
 & 4096    & 1.168e+06   &   98420.0    &   993900.0   &   49310.0    &    6375.0    &   38720.0    &    100     &  0.00949775  &  1.0866e-07  &  0.0431031  \\
 & 8192    & 1.212e+06   &   98380.0    &   999900.0   &   84850.0    &   11020.0    &   69590.0    &    100     &  0.00949819  & 1.80201e-07  &   0.071543  \\
 & 16384   & 1.203e+06   &   98760.0    &  1.011e+06   &   64160.0    &    9772.0    &   50110.0    &    100     &  0.00924395  & 1.86879e-07  &  0.0545578  \\
 & 32768   & 1.218e+06   &   98690.0    &  1.027e+06   &   61290.0    &   12630.0    &   44340.0    &    100     &  0.00818814  & 1.18581e-05  &  0.0515042  \\
 & 65536   & 1.275e+06   &   99420.0    &  1.028e+06   &   111000.0   &   27160.0    &   79410.0    &    100     &  0.00818851  & 7.26453e-06  &   0.089444  \\
 & 131072  & 1.289e+06   &   99150.0    &  1.049e+06   &   97590.0    &   23980.0    &   69150.0    &    100     &  0.00810869  & 5.44987e-06  &  0.0781971  \\
\hline
\rh{Optimized}
 & 1       &  1.03e+06   &   97300.0    &   890000.0   &   23500.0    &    633.0     &   19500.0    &    100     & 5.58992e-06  & 1.54629e-12  &  0.0232673  \\
 & 8       & 1.077e+06   &   97430.0    &   912600.0   &   30030.0    &    898.2     &   25400.0    &    100     & 0.000549208  & 3.68151e-08  &  0.0288196  \\
 & 64      & 1.107e+06   &   97520.0    &   953600.0   &   33230.0    &    1223.0    &   28060.0    &    100     &  0.00763603  & 1.33312e-06  &  0.0305985  \\
 & 512     & 1.128e+06   &   98200.0    &   969400.0   &   35420.0    &    1543.0    &   29790.0    &    100     &  0.0114842   & 7.62847e-05  &  0.0320543  \\
 & 1024    & 1.136e+06   &   98060.0    &   957100.0   &   55370.0    &    4498.0    &   46800.0    &    100     &  0.0114628   & 0.000224632  &  0.0497484  \\
 & 2048    & 1.147e+06   &   98110.0    &   979000.0   &   44080.0    &    2410.0    &   37520.0    &    100     &  0.0112596   & 0.000563262  &   0.039252  \\
 & 4096    & 1.158e+06   &   98480.0    &   995200.0   &   37390.0    &    1984.0    &   31170.0    &    100     &  0.00976431  &  0.00185806  &  0.0330009  \\
 & 8192    & 1.186e+06   &   98510.0    &   999500.0   &   59600.0    &    3978.0    &   51370.0    &    100     &  0.00941109  &  0.0180021   &  0.0513793  \\
 & 16384   & 1.189e+06   &   98760.0    &  1.012e+06   &   48790.0    &    3397.0    &   41100.0    &    100     &  0.0096765   &  0.0436702   &   0.041988  \\
 & 32768   & 1.199e+06   &   98690.0    &  1.028e+06   &   40960.0    &    2596.0    &   34060.0    &    100     &  0.00913887  &   0.282704   &  0.0350085  \\
 & 65536   & 1.234e+06   &   99420.0    &  1.029e+06   &   69350.0    &   10530.0    &   54420.0    &    100     &  0.00911621  &   1.79503    &  0.0577917  \\
 & 131072  & 1.252e+06   &   99160.0    &  1.048e+06   &   60350.0    &    5246.0    &   50630.0    &    100     &  0.0188287   &   5.81354    &   0.049876  \\
\end{tabular}
\caption{4096 cells/core, 2048 directions, hybrid-KBA, PCG max\_it = 10}
\end{sidewaystable}

\begin{sidewaystable}
\centering
\begin{tabular}{c|c|rcrrrrrrrr}
 & \#cores &\ch{Overall} & \ch{Sweep Setup} & \ch{Sweep} & \ch{WG DSA} & \ch{PCG Setup} & \ch{PCG Solve} & \ch{PCG it} & \ch{Resid} & \ch{PCG Resid} & \ch{DSA/Solve}
\\\hline
\rh{Default}
 & 1       &  1.02e+06   &   97300.0    &   893000.0   &   13600.0    &    796.0     &    9390.0    &     42     & 5.59101e-06  & 1.52682e-05  &  0.0134653  \\
 & 8       &  1.06e+06   &   97440.0    &   907300.0   &   18120.0    &    1227.0    &   13180.0    &     46     & 0.000549223  & 3.19653e-05  &   0.017678  \\
 & 64      & 1.095e+06   &   97410.0    &   955000.0   &   20430.0    &    1781.0    &   14720.0    &     46     &  0.00763564  & 2.63186e-05  &  0.0190047  \\
 & 512     & 1.124e+06   &   98120.0    &   975400.0   &   26350.0    &    2741.0    &   19560.0    &     57     &  0.0114609   & 3.91109e-05  &  0.0239111  \\
 & 1024    & 1.134e+06   &   98060.0    &   964300.0   &   46460.0    &    8122.0    &   34290.0    &     53     &  0.0114618   & 8.61561e-05  &  0.0418182  \\
 & 2048    & 1.141e+06   &   98000.0    &   983900.0   &   33280.0    &    4014.0    &   25130.0    &     56     &  0.0110636   & 2.29174e-05  &  0.0297941  \\
 & 4096    & 1.162e+06   &   98400.0    &   999100.0   &   37740.0    &    6383.0    &   27160.0    &     67     &  0.00949776  & 3.56838e-05  &  0.0331926  \\
 & 8192    & 1.195e+06   &   98300.0    &  1.003e+06   &   64700.0    &   11090.0    &   49410.0    &     68     &  0.00949815  & 3.99183e-05  &  0.0553464  \\
 & 16384   & 1.196e+06   &   98670.0    &  1.018e+06   &   50600.0    &    9819.0    &   36520.0    &     69     &  0.00924394  &  6.1145e-05  &  0.0432849  \\
 & 32768   & 1.224e+06   &   98600.0    &  1.033e+06   &   60200.0    &   12630.0    &   43280.0    &     97     &  0.00818815  &  3.6656e-05  &  0.0503766  \\
 & 65536   &  1.28e+06   &   99320.0    &  1.034e+06   &   107700.0   &   27370.0    &   76020.0    &     90     &  0.00818842  & 5.99059e-05  &  0.0865756  \\
 & 131072  & 1.292e+06   &   99140.0    &  1.054e+06   &   90970.0    &   24090.0    &   62440.0    &     90     &  0.0081087   & 6.49262e-05  &  0.0729511  \\ 
\hline
\rh{Optimized}
 & 1       &  1.01e+06   &   97300.0    &   883000.0   &   13900.0    &    633.0     &    9910.0    &     46     & 5.59041e-06  & 1.55471e-05  &  0.0139558  \\
 & 8       & 1.147e+06   &   97400.0    &   991700.0   &   20780.0    &    899.8     &   16160.0    &     60     & 0.000549509  & 7.94732e-05  &  0.0186871  \\
 & 64      & 1.093e+06   &   97490.0    &   945300.0   &   27610.0    &    1221.0    &   22460.0    &     78     &  0.00763566  & 7.50685e-05  &  0.0257556  \\
 & 512     & 1.131e+06   &   98190.0    &   965400.0   &   42970.0    &    1550.0    &   37340.0    &    128     &  0.0114608   & 7.62723e-05  &  0.0387466  \\
 & 1024    &  1.16e+06   &   98090.0    &   954600.0   &   81910.0    &    4405.0    &   73460.0    &    166     &  0.0114619   & 7.20312e-05  &  0.0720405  \\
 & 2048    & 1.169e+06   &   98060.0    &   974000.0   &   71490.0    &    2388.0    &   64960.0    &    181     &  0.0110636   & 6.82923e-05  &  0.0623822  \\
 & 4096    & 1.184e+06   &   98450.0    &   988600.0   &   71060.0    &    1957.0    &   64880.0    &    222     &  0.00949784  & 8.76406e-05  &  0.0612586  \\
 & 8192    & 1.264e+06   &   98410.0    &   994400.0   &   142800.0   &    4130.0    &   134400.0   &    273     &  0.00949827  & 9.77687e-05  &   0.115347  \\
 & 16384   & 1.254e+06   &   98720.0    &  1.007e+06   &   119300.0   &    3190.0    &   111800.0   &    296     &  0.00924396  & 9.87909e-05  &  0.0971498  \\
 & 32768   & 1.273e+06   &   98640.0    &  1.022e+06   &   119600.0   &    2634.0    &   112600.0   &    354     &  0.00818816  & 8.82394e-05  &  0.0962188  \\
 & 65536   & 1.406e+06   &   99230.0    &  1.023e+06   &   245000.0   &   10610.0    &   230000.0   &    454     &  0.00818843  & 8.80992e-05  &   0.178832  \\
 & 131072  &  1.43e+06   &   99130.0    &  1.042e+06   &   240300.0   &    5198.0    &   230600.0   &    492     &  0.00810872  &  9.6962e-05  &   0.173502  \\
\end{tabular}
\caption{4096 cells/core, 2048 directions, hybrid-KBA, PCG resid\_thr = 10$^{-4}$}
\end{sidewaystable}

\begin{sidewaystable}
\centering
\begin{tabular}{c|c|rcrrrrrrrr}
 & \#cores&\ch{Overall} & \ch{Sweep Setup} & \ch{Sweep} & \ch{WG DSA} & \ch{PCG Setup} & \ch{PCG Solve} & \ch{PCG it} & \ch{Resid} & \ch{PCG Resid} & \ch{DSA/Solve}
\\\hline
\rhv{Def. (max\_it)}
 & 1      &  148000.0   &   12900.0    &   129000.0   &    2880.0    &     82.7     &    2380.0    &    100     & 1.07974e-06  & 2.37906e-14  &  0.0198621  \\
 & 8      &  179900.0   &   13020.0    &   159600.0   &    3655.0    &    176.1     &    2911.0    &    100     & 5.58992e-06  & 1.65124e-10  &  0.0206965  \\
 & 64     &  188600.0   &   12880.0    &   166200.0   &    5295.0    &    344.6     &    4271.0    &    100     & 0.000549206  &  7.4149e-09  &  0.0286526  \\
 & 512    &  195700.0   &   12930.0    &   171300.0   &    7017.0    &    767.9     &    5485.0    &    100     &  0.00763565  & 4.17207e-09  &   0.036585  \\
 & 4096   &  206100.0   &   12970.0    &   178700.0   &    9478.0    &    1687.0    &    6944.0    &    100     &  0.0114607   & 1.76634e-08  &  0.0469673  \\
 & 32768  &  228500.0   &   12990.0    &   196300.0   &   13220.0    &    3093.0    &    9186.0    &    100     &  0.00949775  & 1.05777e-07  &  0.0592294  \\
 \hline
\rhv{Def. (resid)}
 & 1       &  147000.0   &   12900.0    &   129000.0   &    1580.0    &     82.7     &    1080.0    &    40     & 1.07993e-06  &  1.0859e-05  &  0.0109722  \\
 & 8       &  179700.0   &   13120.0    &   160500.0   &    2344.0    &    180.6     &    1593.0    &    50     & 5.59006e-06  &  2.2993e-05  &  0.0132955  \\
 & 64      &  188400.0   &   12870.0    &   167600.0   &    3361.0    &    345.9     &    2337.0    &    50     & 0.000549215  & 8.56364e-05  &  0.0182465  \\
 & 512     &  193700.0   &   12910.0    &   171900.0   &    4437.0    &    773.9     &    2893.0    &    48     &  0.00763563  &  4.0598e-05  &  0.0233772  \\
 & 4096    &  204000.0   &   12980.0    &   179500.0   &    6460.0    &    1695.0    &    3913.0    &    52     &  0.0114607   & 7.42235e-05  &  0.0323647  \\
 & 32768   &  226800.0   &   12980.0    &   197400.0   &   10440.0    &    3093.0    &    6403.0    &    67     &  0.00949774  & 2.78747e-05  &  0.0471119  \\
 \hline
 \rhv{Opt. (resid)}
 & 1      &  147000.0   &   12900.0    &   129000.0   &    1570.0    &     69.9     &    1090.0    &     41     & 1.07974e-06  & 1.12727e-05  &  0.0109028  \\
 & 8      &  178700.0   &   12990.0    &   159700.0   &    2340.0    &    140.9     &    1624.0    &     53     & 5.59029e-06  & 2.65561e-05  &  0.0133485  \\
 & 64     &  188000.0   &   12890.0    &   167400.0   &    3553.0    &    267.0     &    2607.0    &     65     & 0.000549322  & 3.53054e-05  &  0.0192784  \\
 & 512    &  193900.0   &   12920.0    &   172100.0   &    4441.0    &    342.6     &    3332.0    &     79     &  0.00763571  & 7.73984e-05  &  0.0233737  \\
 & 4096   &  204800.0   &   13000.0    &   179400.0   &    6621.0    &    434.8     &    5337.0    &    124     &  0.0114608   & 8.91349e-05  &  0.0331547  \\
 & 32768  &  227000.0   &   12960.0    &   197400.0   &   10600.0    &    548.7     &    9107.0    &    205     &  0.00949778  & 6.96932e-05  &  0.0478124  \\
\end{tabular}
\caption{512 cells/core, 2048 directions, volumetric}
\end{sidewaystable}

\begin{sidewaystable}
\centering
\begin{tabular}{c|c|rcrrrrrrrr}
 & \#cores&\ch{Overall} & \ch{Sweep Setup} & \ch{Sweep} & \ch{WG DSA} & \ch{PCG Setup} & \ch{PCG Solve} & \ch{PCG it} & \ch{Resid} & \ch{PCG Resid} & \ch{DSA/Solve}
\\\hline
 \rhv{Def. (max\_it)}
 & 1      &  1.3e+06    &   102000.0   &   1.14e+06   &   24000.0    &    794.0     &   19900.0    &    100     & 5.58992e-06  & 2.96587e-13  &  0.0188976  \\
 & 8      & 1.494e+06   &   102200.0   &  1.335e+06   &   30850.0    &    1224.0    &   25900.0    &    100     & 0.000549206  & 1.79778e-09  &  0.0210007  \\
 & 64     & 1.613e+06   &   101500.0   &   1.45e+06   &   34500.0    &    1676.0    &   28890.0    &    100     &  0.00763565  & 3.41888e-09  &  0.0217254  \\
 & 512    & 1.699e+06   &   101600.0   &  1.533e+06   &   36570.0    &    2244.0    &   30280.0    &    100     &  0.0114607   & 9.63288e-09  &  0.0218589  \\
 & 4096   & 1.714e+06   &   101800.0   &  1.541e+06   &   41130.0    &    3741.0    &   33240.0    &    100     &  0.00949775  & 9.10618e-08  &  0.0243806  \\
 & 32768  & 1.761e+06   &   102100.0   &  1.582e+06   &   47020.0    &    6508.0    &   36240.0    &    100     &  0.00818819  & 1.83731e-05  &  0.0271321  \\
 \hline
 \rhv{Def. (resid)}
 & 1       &  1.29e+06   &   102000.0   &   1.14e+06   &   13500.0    &    794.0     &    9380.0    &    42     & 5.59101e-06  & 1.52682e-05  &    0.0108   \\
 & 8       & 1.486e+06   &   102200.0   &   1.34e+06   &   18130.0    &    1223.0    &   13180.0    &    46     & 0.000549223  & 3.19653e-05  &  0.0124008  \\
 & 64      & 1.599e+06   &   101600.0   &  1.449e+06   &   20630.0    &    1689.0    &   15000.0    &    47     &  0.00763563  & 2.13795e-05  &  0.0131151  \\
 & 512     & 1.684e+06   &   101600.0   &  1.531e+06   &   22820.0    &    2237.0    &   16540.0    &    50     &  0.0114607   & 7.83489e-05  &  0.0137636  \\
 & 4096    & 1.704e+06   &   101900.0   &  1.542e+06   &   31100.0    &    3697.0    &   23260.0    &    67     &  0.00949776  & 2.88819e-05  &   0.018545  \\
 & 32768   & 1.758e+06   &   102100.0   &  1.579e+06   &   46690.0    &    6535.0    &   35880.0    &    99     &  0.00818815  & 5.74289e-05  &  0.0269884  \\
 \hline
 \rhv{Opt. (resid)}
 & 1      &  1.29e+06   &   102000.0   &   1.14e+06   &   13900.0    &    631.0     &    9920.0    &     46     & 5.59041e-06  & 1.55471e-05  &   0.01112   \\
 & 8      & 1.493e+06   &   101900.0   &  1.344e+06   &   20810.0    &    906.7     &   16180.0    &     60     & 0.000549509  & 7.94732e-05  &  0.0141757  \\
 & 64     & 1.608e+06   &   101600.0   &  1.451e+06   &   27520.0    &    1218.0    &   22380.0    &     78     &  0.00763569  & 7.35927e-05  &  0.0173957  \\
 & 512    & 1.698e+06   &   101700.0   &  1.532e+06   &   36090.0    &    1390.0    &   30650.0    &    109     &  0.0114613   & 9.74719e-05  &  0.0215849  \\
 & 4096   &  1.73e+06   &   101800.0   &   1.54e+06   &   57240.0    &    1531.0    &   51560.0    &    188     &  0.0094979   & 9.97187e-05  &   0.033631  \\
 & 32768  &  1.8e+06    &   102200.0   &   1.58e+06   &   87720.0    &    1635.0    &   81810.0    &    302     &  0.00818817  & 7.90693e-05  &  0.0495034  \\
\end{tabular}
\caption{4096 cells/core, 2048 directions, volumetric}
\end{sidewaystable}

\end{document}